\documentclass{article}

\usepackage{arxiv}

\usepackage[utf8]{inputenc} 
\usepackage[T1]{fontenc}    
\usepackage{hyperref}       
\usepackage{url}            
\usepackage{booktabs}       
\usepackage{amsfonts}       
\usepackage{nicefrac}       
\usepackage{microtype}      

\usepackage{xr}
\usepackage{graphicx}
\usepackage{dcolumn}
\usepackage{bm}

\usepackage{amsmath}
\usepackage{amssymb}
\usepackage{mathtools}

\makeatletter
\newcommand*{\addFileDependency}[1]{
  \typeout{(#1)}
  \@addtofilelist{#1}
  \IfFileExists{#1}{}{\typeout{No file #1.}}
}
\makeatother

\newcommand*{\myexternaldocument}[1]{%
    \externaldocument{#1}%
    \addFileDependency{#1.tex}%
    \addFileDependency{#1.aux}%
}
\myexternaldocument{supplement}

\newcommand{\av}[1]{\langle #1 \rangle}

\author{Enrico Ubaldi$^{1}$ \and Raffaella Burioni$^{2,3}$ \and Vittorio Loreto$^{1,4,5}$ \and Francesca Tria$^{4,*}$}
\date{
    $^1$Sony Computer Science Laboratories, 6 Rue Amyot, 75005 Paris, France\\
    $^2$Dept. of Mathematics, Physics and Computer Science, Univ. of Parma, Viale G.P. Usberti 7/A, 43124 Parma, Italy\\
    $^3$INFN, Gruppo Collegato di Parma, Viale G.P. Usberti 7/A, 43124 Parma, Italy\\
    $^4$Sapienza University of Rome, Physics Department, P.le Aldo Moro 5, 00185 Rome, Italy\\
    $^5$Complexity Science Hub Vienna, Josefst\"adter Strasse 39, A-1080 Vienna, Austria\\
    \vspace{.2in}
    $^*$To whom correspondence should be addressed. E-mail: francesca.tria@uniroma1.it\\
    \vspace{.2in}
    \today
}

\title{The exploration of the Adjacent Possible explains the emergence and evolution of social networks}

\begin{document}

\maketitle

\begin{abstract}
The interactions among human beings represent the backbone of our societies. How people interact, establish new connections, and allocate their activities among these links can reveal a lot of our social organization. Despite focused attention by very diverse scientific communities, we still lack a first-principles modeling framework able to account for the birth and evolution of social networks. Here, we tackle this problem by looking at social interactions as a way to explore a very peculiar space, namely the Adjacent Possible space, i.e., the set of individuals we can meet at any given point in time during our lifetime. We leverage on a recent mathematical formalization of the Adjacent Possible space to propose a first-principles theory of social exploration based on simple microscopic rules defining how people get in touch and interact. The new theory predicts both microscopic and macroscopic features of social networks. The most striking feature captured on the microscopic side is the probability for an individual, with already $k$ connections, to acquire a new acquaintance. On the macroscopic side, the model reproduces the main static and dynamic features of social networks: the broad distribution of degree and activities, the average clustering coefficient, and the innovation rate at the global and local levels. The theory is born out in three diverse real-world social networks: the network of mentions between Twitter users, the network of co-authorship of the American Physical Society, and a mobile-phone-calls network.
\end{abstract}

\keywords{Social networks $|$ Adjacent possible space $|$ First-principles modelling $|$ Real-data logging human interactions}

\section{\label{sec:intro}Introduction}

Interactions among individuals shape our current societies and the graph depicting our social interactions can reveal a lot about our social organization and its evolution in time.
That is why social networks have attracted a great deal of attention to understand the mechanisms underlying their evolution and provide valuable information on the microscopic determinants of social dynamics, for instance, individuals' search strategies~\cite{marsili_rise_fall_2004,granovetter1995getting} or the schemes to allocate time in socially charged activities~\cite{Sekara9977,kossinets_empirical_2006}.

The evolution of social networks is shaped by the interplay of diverse and complex mechanisms operating at different scales. Indeed, individuals are likely to engage in social interactions with similar alters ~\cite{davidsen_emergence_small_world_2002,newman_structure_growing_2001,onnela_structure_strength_2007}, for instance connecting to "a friend of a friend" (triadic closure). At the same time, they may seek for novel connections outside of their inner circle of contacts, based on shared interests or experiences (focal closure)~\cite{kumpula_emergenge_weighted_2009,kumpula_community_emergence_2009,kossinets_empirical_2006,granovetter_weak_1977}. 

Social networks are also intrinsically dynamical systems that evolve in time~\cite{holme_temporal_2012,perra2012activity} as links between nodes are continuously created and destroyed~\cite{miritello_limited_2013,miritello_dynamical_2011}. This time-varying nature of the networks deeply affects not only their topological properties~\cite{moinet_burstiness_2015, vestergaard_memory_2014, newman_structure_growing_2001} but also the dynamical processes unfolding on top of their fabrics~\cite{barrat_book_2008,Karsai_weakness_2014,karsai2011small,ubaldi2017burstiness}.

The study and characterization of the complex mechanisms underlying the birth and evolution of social networks have boomed thanks to the growing availability of digital data mirroring human interactions. This circumstance allowed to figure out modeling schemes to capture the essence of many relevant aspects of the whole phenomenology. For instance the propensity of individuals to engage in social interactions~\cite{perra2012activity}, the correlations in the nodes' activity patterns~\cite{barrat_random_itineraries_2013, Karsai_weakness_2014, Ubaldi_asymptotic_2016}, the emergence of topological correlations, such as the assortativity and the clustering of nodes in tightly connected communities linked by bridges~\cite{Laurent_call_2015, cattuto_collective_2009, kumpula_community_emergence_2009}.

Despite being relevant stepping stones, none of the approaches mentioned above is genuinely first-principles. Indeed, they all rely on a set of assumptions, often data-driven. For instance, the distribution of an individual's activity, i.e., the propensity of nodes to engage in social interaction, the tendency to interact with new acquaintances or the mechanism strengthening the old contacts of a node, are typically drawn from empirical measures.

Here we overcome these limitations by proposing a first-principles approach that turns out to be able to explain, without unnecessary assumptions, the birth and evolution of social networks. To this end, we start with the intuition that an exploratory process drives the growth of a social network. In this scheme, individuals expand their circle of acquaintances by exploring a very peculiar space, namely the space of new possible connections. From this perspective, the evolution of a social network is driven by an innovation process through which individuals expand their network of contacts, contributing in this way to the growth of the global social network. 

The above framework is consistent with the notion of {\em Adjacent Possible}~\cite{kauffman1993origins,kauffman_investigations_1996,kauffman_investigations_book_2000}, introduced by the biologist Stuart Kauffman in the framework of molecular and biological evolution. Recently, some of us proposed a mathematical formalization of the notion of the Adjacent Possible~\cite{tria_dynamics_2014,loreto_expanding_2016} where the space of possibilities (for instance represented through a Polya's urn~\cite{polya1930quelques,mahmoud_polya_2008}) grows conditionally to the actual realization of a novelty. In the framework of social interactions the expansion of the Adjacent Possible takes place every time we establish a new connection (link). The individual that we just met gets included in our actual experience and our Adjacent Possible expands to include new nodes that we can potentially met in the future.


Thanks to the possibility of making quantitative predictions through a self-contained mathematical framework, the notion of Adjacent Possible expanded its original scope to encompass studies of innovation processes in human activities~\cite{monechi_waves_2017,iacopini_2018} and technological progress~\cite{saracco2015innovation}. 

The outline of the paper is as follows. The following section summarizes the main stylized facts about the birth and evolution of social networks. Next, we present our first-principles modeling framework. The section devoted to the results offers the set of quantitative predictions drawn from the modeling scheme and their comparison with the collections of empirical data. Finally, we outline our conclusions.

\section*{Stylized facts about social networks}
\label{sec:stylized}

Here, we summarize the empirical data used to test the predictions of our theory. These are three different real-world social networks: {\em i}) the {\bf A}merican {\bf P}hysical {\bf S}ociety (APS) co-authorship network generated by all the papers published in all the APS journals from January 1970 to December 2006.  {\em ii}) The {\bf T}witter {\bf M}ention {\bf N}etwork (TMN) logging all the mentions between users recorded between January and September 2008. {\em iii}) The {\bf M}obile {\bf P}hone {\bf N}etwork (MPN) recording the calls between users of a national provider in an undisclosed European country between January and July 2008. We refer to the methods section and the Section~\ref{sec:si_data} of the Supporting Information (SI) for details. These datasets represent diverse contexts of social interactions, making them an ideal set of empirical observations to test the universality of our model. In particular, the APS dataset describes the undirected interactions of co-authors of a scientific papers ~\cite{newman_structure_scientific_2001, newman_scientific_collab_1_2001, newman_scientific_collab_2_2001, radicchi_scientific_2009}. Here interactions have a high cost in terms of time and resources. The TMN dataset reports the directed citations of a user $i$ citing a user $j$ (that corresponds to an edge from $i$ to $j$) between users of the micro-blogging platform, in which interactions are requiring few resources and can be virtually established from and to any node in the network~\cite{gonccalves2011modeling}. Finally, the MPN dataset lies somewhere in between: communication is not as cheap as in the TMN but still easier than in the APS case~\cite{onnela_structure_strength_2007}. Also, the network may be not single scoped for the users taking part in it: some of them may use it to call close contacts whereas others may use the phone for business reasons~\cite{Ubaldi_asymptotic_2016,miritello_limited_2013}. Let us also note that the TMN and APS datasets account for the growth of the two systems since their onset. Indeed, the effective onset of user adoption for Twitter occurred during 2008~\cite{twit_growth}, whereas the APS created the majority of its journals in 1970. This circumstance ensures a unique test bed for a model of network growth. On the other hand, the MPN situation is more subtle as we have only a limited observation window on a system that underwent a long evolution period beforehand.

\begin{figure*}[t]
\centering
\includegraphics[width=1.0\textwidth]{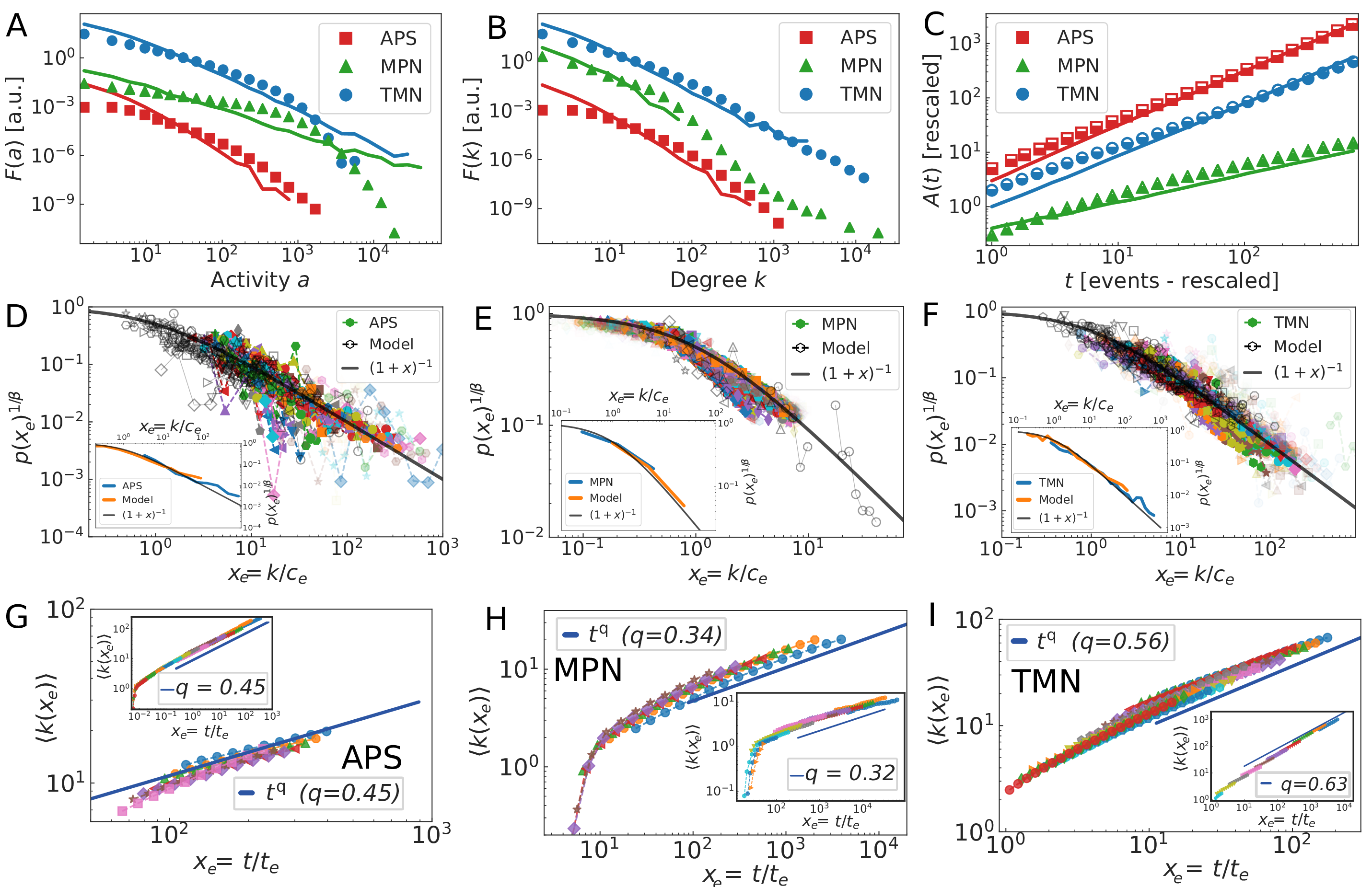}
\caption{
\label{fig:stylized}
    The stylized facts of the three datasets considered in this paper. (A) The $F(a)$ activity distribution, (B) the $F(k)$ degree distribution, and, (C) the temporal growth of the total number of edges $A(t)\propto t^\gamma$ as found in the APS (red squares), MPN (green triangles) and TMN (blue circles) datasets. For each case, we also show the corresponding curves, as found in the model best fitting each dataset (solid lines with the same color as the corresponding dataset). The rescaled strengthening probability $p(k) = (1 + k/c_e)^{-\beta}$ as measured for different classes of nodes (symbols, color depth proportional to the number of agents in a node class $e$) is reported in panel (D) for the APS dataset, in (E) for the MPN case, and, in (F) for the TMN system. In the main panels we compare the empirical curves (colored symbols) with the $p_e(k)$ found in the corresponding best-fitting urn model (black symbols) and the theoretical guideline $p(x_e)^{1/\beta}=(1+x_e)^{-1}$ (black solid lines), being $x_e$ the rescaled degree $x_e=k/c_e$. In the insets we show the average rescaled $\av{p(x_e)}_e$ for the empirical (blue lines) and synthetic data (orange lines) as well as the theoretical behavior $p(x)=(1+x)^{-1}$ (black line). (G-I) Main panels: the average degree $\langle k(t_e,t) \rangle\propto (t/t_e)^q$ as a function of $x_e=t/t_e$ for different classes of nodes entering the system at different times $t_e$ (colored symbols) for the APS (G), MPN (H), and TMN system (I), respectively. In the insets, we show the corresponding results for the urns model. We also show the best fit $\av{k(x_e)}\propto x_e^q$ for all the cases (solid blue lines). Note that here the exponent $q$ does not depend on the class $e$ (refer to SI Section~\ref{sub:si_q_exponent} for details).}
\end{figure*}

\begin{figure*}[t]
\centering
\includegraphics[width=1\textwidth]{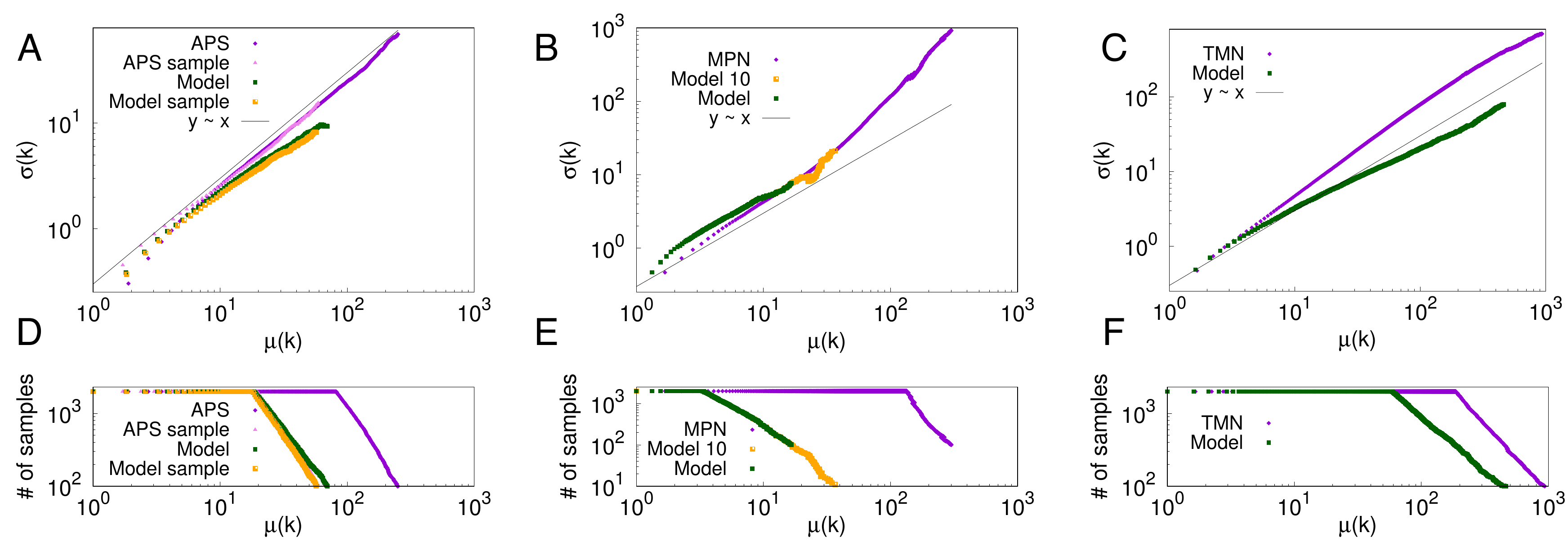}
\caption{
\label{fig:taylor}
    {\bf Top.}   Taylor's law for the number of links created by different individuals at each (intrinsic) time $t$. We defined $\mu(k) \equiv \av{k(t)}$ and $\sigma(k) \equiv \sqrt{\av{k^2(t)} - \av{k(t)}^2}$, with implicit dependence on time. Results in the empirical datasets, contrasted with model predictions, for (A) the APS dataset, (B) the MPN dataset, and (C) the TMN dataset. In Fig. (A) we report results also for the 1-link sampled database, defined in the Methods section. For all the empirical and synthetic datasets we find an approximate power-law behavior  $\sigma(k) \sim \mu(k)^\delta$, with $\delta \gtrsim 1$.  For each curve (both for the empirical and synthetic datasets), we considered the $2000$ individuals with the longest history (final intrinsic time).  {\bf Bottom.} We report here the effective number of points on which the mean and the standard deviation is computed at each time $t$. This number decreases for the highest values of the mean,  corresponding to higher values of intrinsic time since the number of the total performed actions varies from individual to individual. The number of samples used is reported, both for the empirical and synthetic datasets, for (D) the APS dataset, (E) the MPN dataset, and (F) the TMN dataset. In all the cases, we considered only averages on at least $100$ samples, but for the MPN case, where we additionally show values for times where at least $10$ samples were available (Model 10 curves). }
\end{figure*}

The datasets mentioned above have been extensively studied and characterized in previous works~\cite{Ubaldi_asymptotic_2016, Laurent_call_2015, Jo_circadian_2012, Karsai_weakness_2014}, and we resume here their main properties and features. The most renowned property of these systems is that both the propensity of a user to engage in social interaction (i.e., the activity $a_i$ of a node $i$ defined as the number of events actively engaged by node $i$) and the degree $k_i$ (i.e., the number of different neighbors connected to node $i$) are found to be broadly distributed. The tails of their distributions are usually approximated with a power-law, i.e.,
 $$
 F(a)\propto a^{-(\eta+1)} \quad \text{and} \quad F(k)\propto k^{-\mu},
 $$ 
 as shown in Fig.~\ref{fig:stylized}A-B.

These systems are also expanding in time as new nodes and edges keep entering the network. In Fig.~\ref{fig:stylized}C we show the growth in intrinsic time $t$ (i.e., the number of recorded events) of the number of edges $A(t)$ in the systems  that follows a Heaps' law as 
$$
A(t)\propto t^\gamma
$$ 
(see  Section~\ref{sec:si_results} in the SI for details).

Another key feature of social interactions is that individuals display correlations on their activity. When a node engages a social interaction, it is likely to turn its social activity (e.g., a mention in the TMN) toward a node already contacted in the past rather than toward a randomly selected node in the system. A possible way to quantify this mechanism is to measure the probability $p_{k\to k+1}(k)$ (in short $p(k)$) for a node that already contacted $k$ different nodes to contact a new one the next time it will be active~\cite{Karsai_weakness_2014,Ubaldi_asymptotic_2016}. The $p(k)$, which is formally the probability to pass from degree $k \to k+1$, was found to feature the same functional form
 $$
 p(k) = \left(1+\frac{k}{c}\right)^{-\beta}
 $$
 across all the datasets we here consider~\cite{Ubaldi_asymptotic_2016}. Specifically, it turns out that we can characterize a system through a single value of $\beta$ (the strengthening exponent) and a distribution of values of the strengthening constants $c$. The constant $c$ sets the scale at which an individual decreases his ability to acquire new contacts. At odds with the strengthening exponent $\beta$, $c$ significantly varies across individuals. To further explore this variability, we grouped individuals in different classes, $e$, according to their entrance time $t_e$ into the system and their final degree $k_e$ (refer to the Methods section). We show in Fig.~\ref{fig:stylized}D-F both the $p(x_e)$, with $x_e=k/c_e$ for each class $e$, (main panels) and the average value of the rescaled  probability $\av{p(x_e)}_e$ (insets), as found in the empirical data and, for comparison, in the  synthetic results of our modeling framework (see also methods and SI Section~\ref{sec:si_results} for details).  This tendency to slow down the creation of new links also reflects in a sub-linear growth of the average degree  
 $$
 \av{k(t_e,t)}\propto t^q,
 $$
  with $q<1$, as shown in Fig.~\ref{fig:stylized}G-I. 
We further explore for the first time a  finer-grained behavior, Taylor's law~\cite{taylor_1961,eisler2008},  recently pointed out as a shared feature in evolving systems, when referred to fluctuations in innovation rate~\cite{gerlach2014,tria2020}. Taylor's law relates the standard deviation of a random variable to its mean, and the onset of complex behavior is characterized by
 $$
 \sigma(t) \propto \mu(t)^\delta
 $$
  with $\delta\geq 1$,
at odds with $\delta=1/2$ characterizing uncorrelated events. A complex behavior of Taylor's law does not trivially follow from Zipf's and Heaps' laws~\cite{gerlach2014,tria2020},  making it a relevant observable to test theoretical predictions.  We here measure Taylor's law referred to the growth of individual connections in the network, observing a linear or superlinear behavior (Fig.~\ref{fig:taylor}).

In addition to these global observables, we also track a set of local observables, later presented in the Results section, measuring how agents allocate their events in their local network of contacts, e.g., reinforcing old contacts or establishing new links either closing or not closing existing triangles in the emerging social graph. 



In the following, we propose a minimalistic, first-principles model of network evolution that reproduces all the features mentioned above and provides more in-depth insight into the microscopic dynamics shaping the growth and evolution of social networks.

\section*{A first-principles model for social networks}
\label{sec:model}

We now introduce our first-principles model for the birth and evolution of social networks. The model we propose builds on the expansion of the Adjacent Possible framework~\cite{tria_dynamics_2014} to the exploration of social spaces where individuals are embedded. In this framework, we can microscopically model the space of possibilities of a node, i.e., the set of all the social interactions that are "possible" for a node within the social network. This space, at a given point in time, consists of three distinct regions: i) the {\em actual}, including all the links already experienced by the individuals in the past (current connections),  ii) the {\em adjacent possible} comprehending all the links that are just one step away from being explored (e.g., the friends of friends that we still do not know), and, iii) the {\em unknown}, accounting for all the links not yet conceivable by the node at present, but that may become adjacent and possible at some later stage.


A second essential ingredient of our modeling scheme concerns the phenomenon of the so-called correlated novelties~\cite{tria_dynamics_2014}. Every time the social exploration process of a node $i$ activates a new connection with a node $j$ belonging to its adjacent possible, $i$ and $j$ experience a novelty, i.e., the link $e_{ij}$ gets active for the first time. In this way, $j$ becomes now part of the {\em actual} region of $i$ and the {\em adjacent possible} of $i$ reacts to this novelty by surrounding it with freshly created adjacent possible, i.e., new possible connections that were not possible for $i$ before. In other words, a novelty paves the way to another in the future.

\subsection*{Model rules} 
\label{sub:model_definition}

The modeling scheme we apply is a multi-agent version of a modified Polya urn~\cite{polya1930quelques, mahmoud_polya_2008} that proved to be able to reproduce the adjacent possible evolution in different contexts.  In particular, it was proven to account for the main statistical features shared by complex evolving systems (Zipf's, Heaps' and Taylor's law) and observed correlations in a more sophisticated version~\cite{tria_dynamics_2014,monechi_waves_2017,entropy_2018,tria2020}. In the simpler formulation of that model~\cite{tria_dynamics_2014}, the key ingredient is an urn, $\mathcal{U}$, containing $N_0$ distinct elements. One may think of them as balls of different colors representing an item of the space being explored. The dynamics proceeds by repeatedly withdraw balls from $\mathcal{U}$ and annotating them in a temporal sequence of events (this sequence may alternatively represent, depending on the context, a sequence of phone calls made, or a list of co-authors of new scientific publications or a sequence of retweets). Every time we pick up a ball, we put it back in the urn together with $\rho$ additional copies of it, thereby reinforcing that element's likelihood of being drawn again in the future, in a ``rich-get-richer'' fashion. Also, to account for the adjacent possible expansion, whenever a novel (never extracted before) element appears in the sequence, we additionally put $\nu+1$ new distinct elements in $\mathcal{U}$, thus expanding the adjacent possible of the system.

\begin{figure*}[t]
    \centering
    \includegraphics[width=.7\textwidth]{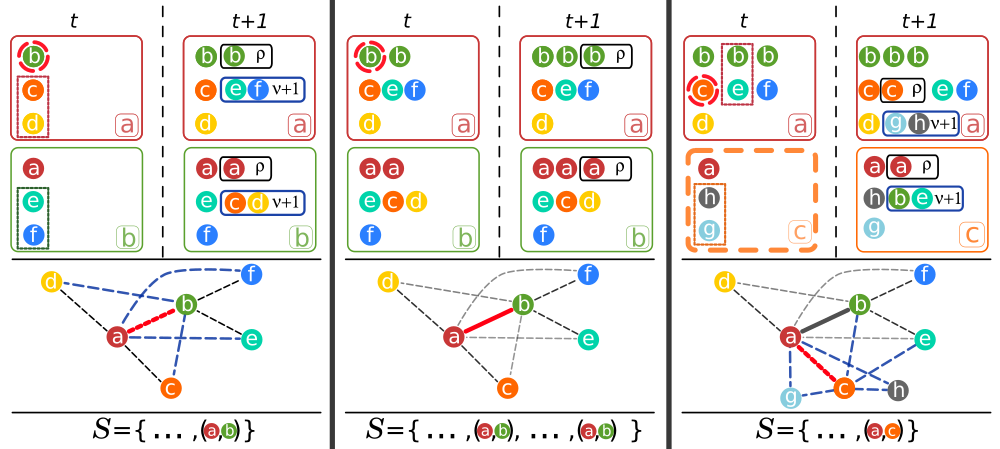}
    \caption{
        \label{fig:model} Three possible steps of the Polya's urn model for a system with $\rho = \nu = 1$ with sampling strategy $s=$WSW. For each evolutionary step of the system (columns), we show the current state of the urns (top row), the equivalent network evolution (mid row), and, the sequence $\mathcal{S}$ of observed events (bottom row). In the network, we show already active links (solid lines), links in the adjacent possible (dashed lines), currently active links (red lines) and connections entering into the adjacent possible (blue dashed lines). {\bf First column}: at time $t$ urn $a$ is active and draws a ball with ID $b$ (red-circled ball): the event $(a,b)$ is then appended to the sequence $\mathcal{S}$. At time $t+1$ the $a$ urns then gains $\rho$ copies of $b$ and vice-versa (reinforcement) and, since the $e_{ab}$ link is new, we also draw $\nu+1$ distinct balls from $a$ following the WSW strategy (balls $c$ and $d$ within dashed rectangle) that will be copied into $b$ (and the same for $b$ that sends $e$ and $f$ as novelties to $a$). Mid row: the $e_{ab}$ edge is active and the $e_{ae},\,e_{af},\,e_{bc}$ and $e_{bd}$ links enters into the adjacent possible. Notice that the adjacent possible of $c$ changed without the need for $c$ to participate in a social interaction. {\bf Second column}: at time $t$, urn $a$ draws a copy of $b$ (top). Since the edge $e_{ab}$ was already active in the past, we only put $\rho$ copies of $b$ in urn $a$ and the other way around. The network's topology does not change in this step,  while the weight of the $e_{ab}$ link gets increased (mid row). \textbf{Third column}: urn $a$ draws a copy of $c$ at time $t$ (top). Since $c$ is an empty urn, it creates $\nu+1$ novel IDs ($g$ and $h$, in the dashed rectangle) and gains a copy of them. We add $(a,c)$ to the sequence $\mathcal{S}$ and we perform the reinforcement/novelties exchange between $a$ and $c$. The network gains two new nodes ($g$ and $h$), activates a new edge ($e_{ac}$) and inserts new links in the adjacent possible. The actual space of $c$ acquires $a$ while its adjacent possible gains $e,\,g$, and $h$.
        }
\end{figure*}

To account for the birth and evolution of social networks, we generalize this model to a multi-agent version. The paradigmatic shift is twofold. On the one hand, the system will consist of a collection of urns, each identified by a unique alphanumeric ID ($a, b, c, \ldots$), representing users in a social network. On the other hand, each ball within each urn will bear the reference ID of another urn in the system. Then, the sequence of extracted balls will be the series of social contacts annotated as tuples $(i,j)$, where $i$ is the ID of the urn drawing a ball, and $j$ is the ID of the drawn ball. For each extraction, the reinforcement process requires to put back $\rho$ copies of the extracted ball $j$ into the extracting urn $i$ (and vice versa), so that an exploited interaction will be favored again in the future. To account for the expansion of the adjacent possible, we also let two urns that interact for the first time to exchange a {\em memory buffer}, i.e., a subset of $\nu+1$ balls each urn that is reciprocally shared.  Thanks to this exchange, an urn that experiences a novelty, i.e., that establishes a connection never explored before, expands its adjacent possible ---i.e., the set of IDs that it may contact in the future. A schematic representation of the model is given in Fig.~\ref{fig:model} and we resume here the steps defining it (see the methods section and the SI Section~\ref{sec:si_model} for further details):

\begin{enumerate}
    \item[1)] we start with two urns, $a$ and $b$ having a copy of each other's ID inside of them; the urns also contain the $\nu+1$ distinct identities (IDs) of other urns that did not participate yet to any interaction ($c$, $d$ for $a$ and $e$, $f$ for $b$).  This set is their  {\em memory buffer}  at the initial stage.  We shall come back later on the different strategies to update it during the evolution of the system. The sequence of events $\mathcal{S}$ is initially empty;
    \item[2)] at each time step, we extract a ``calling'' urn $i$ with probability proportional to the size of the urn $U_i$ (the number of balls within the urn $i$). We then draw a ball from the calling urn $i$, say the ID $j$. This double extraction corresponds to a single event $(i,j)$ that we append to the main sequence $\mathcal{S}$. In Fig.~\ref{fig:model} the first event is the $(a,b)$ one.
    \item[3)] reinforcement: following the event $(i,j)$, we add $\rho$ copies of $i$ in the $j$'s urn and $\rho$ copies of $j$ in the $i$'s urn. For example, in the first column of Fig.~\ref{fig:model}, we add $\rho$ copies of $a$ in the $b$'s urn and $\rho$ copies of $b$ in the $a$'s urn.
    \item[4)] novelty: if it is the first time that $i$ and $j$ interact, $i$ and $j$ exchange their {\em memory buffer}. With this mechanism, we add $j$'s memory buffer into $U_i$ and, vice-versa, $i$'s memory buffer into $U_j$. In Fig.~\ref{fig:model}, first column, $a$'s memory buffer ($c$, $d$) is copied into $U_b$ and  $b$'s memory buffer ($e$, $f$) is copied into $U_a$.
    \item[5)] if a node $j$ is called for the first time by another node (i.e., $j$ is an empty urn so that $U_j=0$), it creates $\nu+1$ new agents (empty urns) and, for each of them it creates a ball into its urn: these $\nu+1$ IDs represent the initial memory buffer of $j$. In Fig.~\ref{fig:model} (third column) node $c$ creates two brand new nodes, $g$ and $h$, that will represent its initial memory buffer. We note here that the newly created agents are initially empty urns so that they can participate in the dynamics (they can be included in the social network) only if another urn (agent) calls them. Only after this first call, they become active. In this scheme, an agent cannot join the network "from outside," i.e.,  unless it is engaged by another agent already belonging to the network. Of course, the scheme can be generalized to include a sort of "immigration," but this aspect will not be discussed in the present work.
\end{enumerate}
Each evolution step is defined as a repetition of the $2\to 5$ steps of the just outlined procedure, as shown in Fig.~\ref{fig:model}. 
The parameters $\rho$ and $\nu$ weigh the relative importance of the reinforcement and exploration processes in the system. We define $R=\rho/\nu$ as the ratio between the two.

\subsection*{Sharing of past experiences} 
\label{sub:strategies}

The last ingredient of our modeling scheme is the strategy that an agent adopts when sharing its experience (the memory buffer) with nodes encountered for the first time. To this end, we introduce different strategies $s$ to determine the $\nu+1$ IDs contained in the memory buffer being shared along with new links. Here, we report three of these strategies that turn out to best capture the phenomenology of the empirical datasets we consider, while we refer to the SI Section~\ref{sec:si_model} for additional strategies. (i) {\bf W}eighted {\bf S}ample with {\bf W}ithdrawal (WSW) strategy: an agent draws $\nu+1$ {\em distinct} IDs from the urn proportionally to their abundance in the urn itself at that time, i.e., proportional to the number of the past interactions with each ID. This strategy corresponds to sharing the IDs that interacted the most with a node in the past and is the one applied in Fig.~\ref{fig:model}; (ii) {\bf S}ymmetric {\bf S}liding {\bf W}indow (SSW): each agent keeps a buffer of its last $\nu+1$ interactions that represent the list of IDs shared with a newly contacted agent. After the exchange, both agents update their memory buffer by pushing in the ID of the agent just contacted and removing the $\nu+1$-th ID from their buffers. This strategy favors the spreading in the network of the recently activated connections, rather than the most frequent ones; (iii) {\bf A}symmetric {\bf S}liding {\bf W}indow (ASW): it is a variant of the previous one, where only the agent that initiated the interaction updates its memory buffer after the communication event.

The model is then entirely defined by three parameters only: the reinforcement value $\rho$, the ratio $R=\rho/\nu$ setting the relative importance of the reinforcement (exploit) and novelties (explore) mechanisms, and strategy $s$ used to exchange the memory buffer between nodes getting in contact for the first time.

\section*{Results}
\label{sec:results}

In this section, we report the main features emerging from our modeling scheme's evolution and compare them to the empirical data of the three real-life social networks we previously introduced. To better compare the theoretical predictions with the observed data, we optimized the model by fixing the parameter values to maximize, for each empirical dataset, a score function $S_d(\rho, R, s)$.  It evaluates the goodness of fit of the synthetic simulations to the empirical dataset $d$ by looking at eight selected observables, both local and global, static and dynamic (see the methods section for details). 
We consider in particular the exponents of the power laws discussed in section~\ref{sec:stylized}, as well as topological measures: (i) the average clustering coefficient, (ii) the fraction of newly created edges that are either old (already activated in the past), new (being activated now) and that happen to close a triangle (closed) or not (open). We create thus four categories: the old open (OO), the old closed (OC), new open (NO) and new closed (NC), and measure the fraction of events falling in each category per time range in the asymptotic limit of the system evolution (see also the methods section).
We summarize the results in radar plots in Fig.~\ref{fig:results}, showing the observed values for the eight selected observables along with the best theoretical estimates. 
We observe that the model endowed with optimal values of the parameters is able to reproduce all the selected observables in all the datasets quantitatively, but the APS dataset, where the model fails to predict the observables related to the network topology correctly. To explain this discrepancy, we note that the APS dataset is composed of cliques of events ---rather than one single event between two ids per time. 
This feature leads to a naturally significant clustering coefficient (as all the agents publishing one paper are fully connected) and in an increased count of events observed along old edges insisting on at least one closed triangle.  To filter out this effect, we performed a sub-sampling of the data by drawing a single link among all the possible ones for each paper and re-computed the features of this sub-sampled dataset (see the methods section and the SI Section~\ref{sec:si_APS_subsample} for further sampling strategies). In this way, the disagreement between the real system and the model results disappears  (see Fig.~\ref{fig:results}(A)), revealing that the model can explain the underlying interaction processes also in this dataset.

In the SI, we additionally show how the model parameters affect the main observables of the system. Here we say that, as for the global observables discussed in section~\ref{sec:stylized} and reported in Fig.~\ref{fig:stylized} and Fig.~\ref{fig:taylor}, the relevant parameters are the ratio $R=\rho/\nu$ and the sharing strategy $s$, while the absolute values of $\rho$ and $\nu$ impact the behavior of the observables related to the local topology of the network. Despite the limited number of parameters, the model is flexible enough to reproduce a wide range of phenomenologies, from highly exploratory situations (high $\gamma$ and $\beta$ exponents) to more exploitative scenarios, with a reduced number of connections being explored.

\subsection*{Global trends}
\label{sec:global_trends}
To make contact with real-world social networks, in Fig.~\ref{fig:stylized}, we contrast the synthetic results with the empirical datasets considered. Specifically, we show that the model can reproduce broad activity (A) and degree distributions (B), the time evolution of the number of edges in the network $E(t)\propto t^\gamma$ (C), the functional form of the strengthening function $p(k)$ (D-F), as well as the sub-linear growth in of the average degree (G-I). Further, the model can reproduce the broad fluctuations in the rate of innovation (i.e., in the creation of new links), as measured by Taylor's law (Fig.~\ref{fig:taylor}). All these observables are discussed in section~\ref{sec:stylized}.

\begin{figure*}
    \centering
    \includegraphics[width=.9\textwidth]{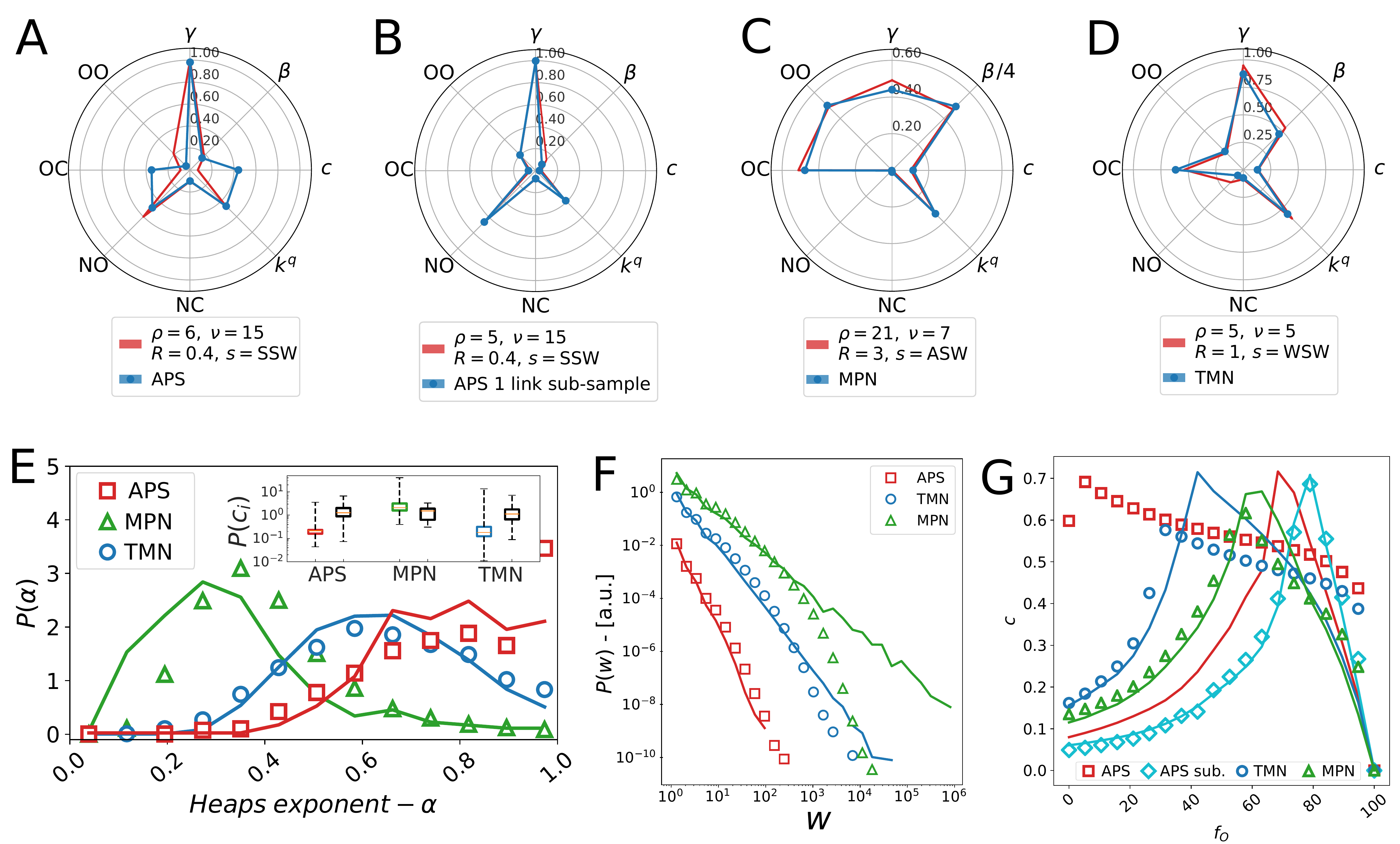}
    \caption{
        \label{fig:results} (A-D) Radar plots comparing eight selected observables measured in empirical and synthetic data. For each dataset, we show the observed values (blue lines) and the corresponding values in the synthetic model fitting (red lines). For each fitting model, we report the reinforcement and novelties parameters $\rho$ and $\nu$ and their ratio $R$ as well as the optimal sampling strategy $s$.  (E) The distribution of the local Heaps' exponent $\alpha$ for a sample of $10\%$ of nodes as measured in the empirical datasets (symbols) and the synthetic simulations (solid lines) for the APS (red), MPN (green), and TMN (blue) datasets. In the inset, we show the distribution of the strengthening constant $c_i$ as measured in data (colored boxes) and corresponding simulations (black boxes). (F) The $P(w)$ link weight distribution for the three empirical datasets (symbols) and the ones found in the artificial networks (solid lines). (G) The average clustering coefficient $c(f_O)$ as a function of the fraction of removed edges with overlap $O\le f_O$. Symbols refer to empirical data, while solid lines represent the synthetic results (the cyan color is used for the $1$-link sampling of APS). 
    }
\end{figure*}

We summarize the results in radar plots in Fig.~\ref{fig:results}, showing the observed values for the eight selected observables along with the best theoretical estimates. With these parameters' values, we compare the model predictions with empirical findings for different local observables.

\subsection*{Heterogeneities in the experience of the new}

Interestingly, the model also correctly captures the heterogeneous propensity of individuals to establish new connections, i.e., the rate at which they experience novelties. To quantify this rate, we look at the exponent of the Heaps' law describing the growth of the degree of an individual, $k_i(x_i)$, i.e., the number of distinct people encountered as a function of the number of social events performed $x_i$: $k_i(x_i) \propto x_i^\alpha$. Fig.~\ref{fig:results}(E) reports the distribution of empirical exponents $\alpha$ for the three datasets considered. These distributions are peaked at different $\bar\alpha_d$ values for the different datasets ($\bar\alpha_{\rm APS}\sim 0.9$, while $\bar\alpha_{\rm TMN}\sim 0.7$ and $\bar\alpha_{\rm MPN}\sim 0.4$). We also report the $P(\alpha)$ distributions as obtained using our modeling scheme. Remarkably, the model correctly reproduces both the peak value and the broadness of each empirical $P(\alpha)$ distribution. 

Another signature of heterogeneity in the empirical data is represented by the distribution of the strengthening constants $c_i$. We remind that the constants $c_i$ enter the probability for an individual with $k$ connections to acquire a new one:
 $$
 p_i(k) = (1+k/c_i)^{-\beta},
 $$
\noindent where the coefficient $c_i$ modulates the propensity of individual $i$ to create new connections. The inset of Fig.~\ref{fig:results}E illustrates how our modeling scheme qualitatively reproduces the empirical $P(c_i)$ of the strengthening constants $c_i$. This is another important result, already anticipated in Fig.~\ref{fig:stylized}, as the model synthetically reproduces the different propensity of individuals in a social network to decrease their social exploration at a given cumulative $k$.

\begin{figure*}
	\includegraphics[width=1\textwidth]{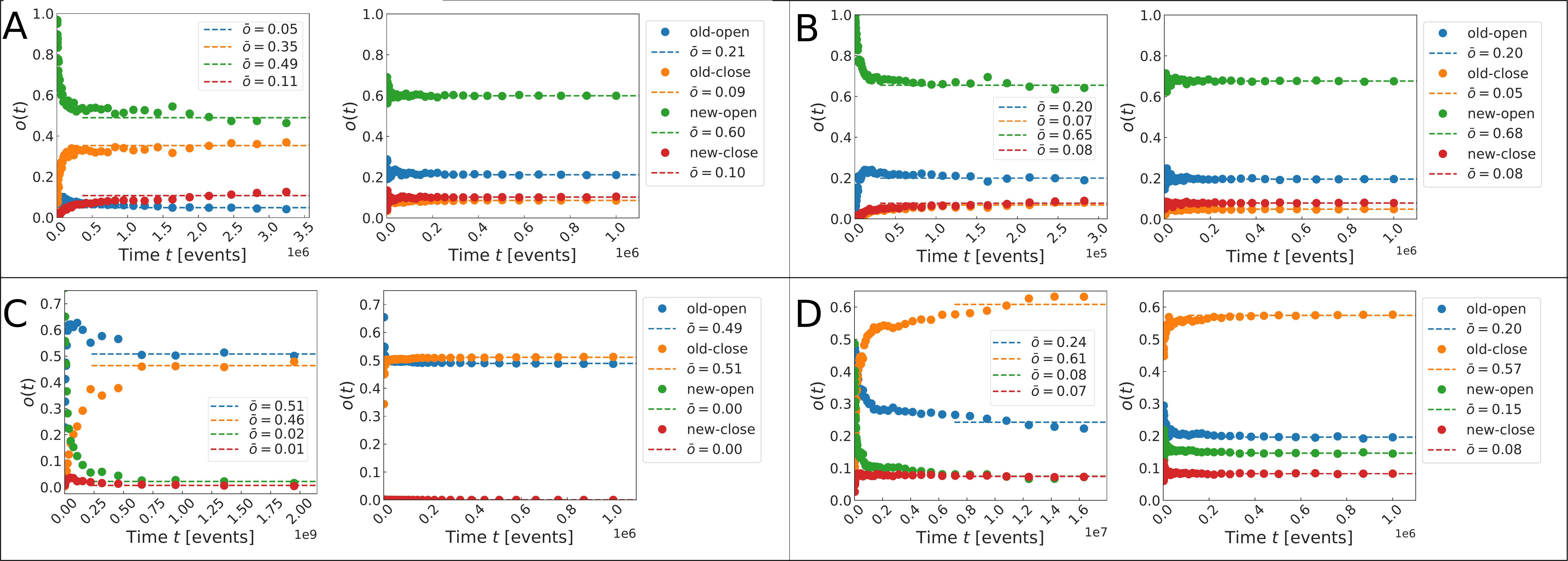}
    \caption{
        \label{fig:openClose}
        Plot of the empirical data (left column in each panel) and model results (right column in each panel) results for $OO(t)$ (blue markers), $OC(t)$ (orange markers), $NO(t)$ (green markers), and $NC(t)$ (red markers). In each case, we show the temporal behavior (markers), and the asymptotic value (dashed lines, the asymptotic value reported in the legends on the right of the panel). Panels refer to (A) APS, (B) $1$-link subsampling of APS, (C) the MPN dataset, and (D)  the TMN dataset.
    }
\end{figure*}

\subsection*{Topological correlations}

We further expand the comparison between empirical and synthetic data checking the topological correlations of the weighted network of interactions among individuals. First, the model correctly reproduces the overall link weight distribution $P(w_{ij})$, i.e., the distribution of the number of activations of a single edge $w_{ij}$ (Fig.~\ref{fig:results}F). We finally observe that both the empirical and the synthetic data obey the weak and strong ties scheme of the Granovetter conjecture~\cite{granovetter1995getting, granovetter_weak_1977, Laurent_call_2015}. The latter states that links in a social system will be arranged to have communities of individuals tightly connected by strong ties and with a large neighbors overlap. These communities are then interacting through weak ties, i.e., links acting as bridges between communities between nodes sharing a limited number of common neighbors (low overlap). To prove this, we measure how the average clustering coefficient $c(f_O)$ of the network varies by removing edges by their ascending overlap $O_{ij}$, i.e., the fraction of common neighbors between nodes $i$ and $j$ (see Methods section for details).  In Fig.~\ref{fig:results}(G), we plot the average clustering coefficient $c(f_O)$ computed in a network where we remove all the edges $e_{ij}$ with overlap $O_{ij}\le f_O$, being $f_O$ the percentile of the overlap distribution. We find $c(f_O)$ to increase as one removes edges with small overlap (indicating that the removal of weak ties is removing bridges between communities) until the $c(f_O)$ peaks. After the maximum, if we keep removing the higher overlap edges, we start breaking the triangles in the communities' cores, and the clustering coefficient decreases. The APS dataset is found to be in disagreement with synthetic data. As already noted above, this disagreement can be explained, remembering that in these datasets, events are composed of cliques of interacting authors, whereas the model only accounts for pairwise interactions. To filter out this difference, we repeated the clustering measure vs. overlap curve on the $1$-link sub-sampled APS dataset. The result is reported in Fig.~\ref{fig:results}(G), showing a perfect agreement between the model and the data. 

As a last measure to investigate the agreement between the model predictions and empirical findings on the microscopic dynamics, we report the temporal evolution of the number of events allocating toward new or old links insisting or not on triangles (the categories $OO$, $OC$, $NO$, and $NC$ defined above and in the methods section). Again, we note that, while the original dataset APS shows a behavior not well reproduced by the model, the latter very nicely predicts the behavior featured by the $1$-link sub-sampled APS dataset, as well as the behavior shown by the other two datasets (Fig.~\ref{fig:openClose}).


\subsection*{Exploration strategies}

 Let us finally note that the parameter configuration found to describe better each dataset draws some meaningful insights on the microscopic mechanisms driving the exploration of the social space at the individual level. In the TMN case, we find $R=1$, so that the reinforcement and the novelty exchange processes equally influence the single agents' exploration process: this is reasonable in a system where new connections require little effort from the user. Moreover, the strategy $s=$WSW with $\nu=5$ is the one that better describes the empirical data: users select new accounts to follow by sampling from the past interactions of the alters they are connecting with proportionally to their popularity.

On the other hand, in the MPN case, the best fit is obtained for $R=3$ and $\rho=21$. The system dynamic is dominated by reinforcement processes that tend to reinforce links that are first established and inhibiting the creation of new edges. In this case, the best fit with the $s=$ASW memory buffer sampling strategy highlights that individuals share their last $\nu+1=8$ contacts, thus spreading copies of recently contacted IDs rather than the most contacted ones. Notice that the last contacted $\nu+1$ IDs may, in general, be different from the most representative IDs within the urn. The asymmetric nature of the ASW strategy indicates that users actively exploring new connections update their memory buffer, whereas nodes passively participating in communication tend to conserve their previous memory buffers.

Finally, in the APS case, we find an extremely exploratory dynamics characterized by a relatively low $R=0.4$, i.e., a relatively high $\nu$. This finding is symptomatic of a dynamics where the exploration of the social space overtakes the reinforcement of existing connections. A possible explanation lies in many students and researchers authoring a few papers before quitting academia and research, providing a constant influx of new potential connections to be explored by senior researchers. The $SSW$ optimal sampling strategy reveals that authors tend to share their last $\nu+1 \simeq 16$ people they have been collaborating with, implying a preference to recommend recently active connections to new collaborators. Moreover, this strategy also catches the intrinsic symmetric nature of the co-authorship interaction, as both co-authors update their buffers of potential new collaborators.

\section*{Discussion}

In this work, we proposed a first-principles theoretical model of social exploration to explain the birth and evolution of social networks. The theory is based on the notion of Adjacent Possible and builds on a recently introduced mathematical formalization of its conditional expansion. In this framework, the creation of new social bonds is the outcome of an exploration process unfolding on the space of possible new acquaintances, whose boundaries change while people explore them. 

Without relying on unnecessary assumptions, our new theory starts from first principles and predicts both microscopic and macroscopic features of real-world social networks. We compared the predictions with the empirical data from three diverse social networks: the system of Twitter users, the network of co-authorship of the American Physical Society, and the phone-call network. The agreement between theory and data is surprisingly good. On the macroscopic side, the model reproduces the main static and dynamic features of those social networks: the broad distribution of degree and activities, the average clustering coefficient, and the innovation rate at the global and local levels. At the microscopic level, the most striking feature captured is the probability for an individual, with already $k$ connections in its local network, to acquire a new acquaintance. The model also captures topological correlations and the dynamics of real-world systems at very different scales, from the local exploit/explore mechanisms of single agents to the global organization of the network in communities of coherent users. 

Besides being able to capture very complex features of social networks quantitatively, our theory also allows us to deepen our understanding of the microscopic mechanisms shaping the propensity of people to reinforce old contacts or establish new ones. For instance, in Twitter mentions network, we find the exploration and reinforcement processes to be of equal importance. Moreover, when getting in contact with new alters, users share a sample of their most common contacts as new potential connections. On the other end of the spectrum, in the mobile phone-calls network, people reinforce their existing bonds more than explore new ones. When suggesting new potential contacts to others, people tend to exchange their most recent contacts, rather than their most common ones. Finally, the network of scientific co-authorship of the American Physical Society journals features the most exploratory dynamics, with new connections massively expanding the adjacent possible of a single node. In this case, people preferably share their last contacts, and the optimal synthetic update procedure is symmetrical, correctly reproducing the intrinsic symmetric nature of the interactions.


The theoretical framework proposed here is, of course, open to possible improvements. First, the simulated dynamics describes the evolution of a system from its outset. The initial conditions set here could be far from those of the real-world systems considered. Despite the excellent agreement with empirical data, a more comprehensive study on the dependence of the system evolution on the initial state is in order. Other generalizations could concern the possibility to remove links or to decouple the rate of addition of links from the that of the entrance of new nodes. Finally, our modeling scheme does not account for effects connected to semantics or affinity between people. For instance, it seems reasonable to assume that people create bonds and interact based on shared interests or their level of homophily. The generality of the approach presented here will make the extension of the theoretical framework desirable and possible along these lines.

Nevertheless, we believe that the presented framework, together with its predictions validated on real-world social networks, represents a fundamental step toward understanding the processes underlying the birth and evolution of social networks. It further creates an important bridge between network theory and urns models, opening the way to constructive contamination between the two fields and a full exploitation
of results derived for stochastic processes relevant in innovation dynamics~\cite{pitman,entropy_2018,tria2020}.
This development, in turn, unlocks the possibility to grasp the very essence of social interactions and allows for the design of efficient and informed policies to address crucial challenges dealing with collective processes ongoing in social networks, such as the spread of diseases and online misinformation.


\subsection*{Contributions} 
\label{ssub:Contributions}
All authors conceived and designed the research work. EU ran the simulations and analyzed the data. All authors wrote and reviewed the article.

\subsection*{Author declarations} 
\label{ssub:Declarations}
The authors declare no conflict of interest.

\subsubsection*{Acknowledgement} 
\label{ssub:subsubsection name}
The authors would like to thank M. Karsai for useful comments and for granting us access to the mobile phone dataset and V.D.P. Servedio for inspiring discussions and relevant suggestions.

\subsection*{Materials and Methods}
\subsubsection*{Data and code}
The three datasets used in the study are:
\begin{itemize}
  \item The co-authorship networks found in the Journals of the American Physical Society~\cite{radicchi_scientific_2009} covering the period between Jan. $1970$ to Dec. $2006$ and containing $301,236$ papers written by $184,583$ authors that are connected by $995,904$ edges.
  \item Twitter Mentions Network (TMN), containing all the mention events exchanged by users from January to September $2008$. The network has $536,210$ nodes performing about $160$M events and connected by $2.6$M edges;
  \item Mobile Phone calls Network (MPN) composed of $6,779,063$ users of a single operator with about $20$\% market share in an undisclosed European country from January to July $2008$. The datasets contain all the phone calls to and from company users, thus including the calls towards or from $33,160,589$ users in the country connected by $92,784,825$ edges.
  \item The synthetic simulations have been run for $T=10^6$ evolution steps for configurations with $R \le 1$, $T=5 \cdot 10^7$ otherwise.
\end{itemize} 
The code used to run the simulations, all the analysis code, as well as the synthetic data analyzed,  are available in~\cite{github_urns}. Due to data policies and IPR, we cannot share the MPN and TMN data, while the APS data are from the work in~\cite{radicchi_scientific_2009}.

\subsubsection*{Asymptotic behavior of the system}
In this work, we leverage on a previous analysis performed on the same datasets as found in ~\cite{Ubaldi_asymptotic_2016}. Specifically, we measure the strengthening probability $p(k)$, i.e., the probability for an individual who already contacted $k$ distinct individuals in the past to contact a new one (i.e., a new node of the network). To average this probability on homogeneous classes of people, we divide the nodes in $g=1,\ldots,G$ classes depending on their time of entrance in the system, $t_e$, and their final degree $k_e$, one class for each combination of $t_e$ and $k_e$. The functional form of the probability $p(k)$ is found to depend on the class $e$ of the nodes as $p(k_e)=(1+k/c_e)^{-\beta}$ with a single overall $\beta$ exponent and a distributed reinforcement constant $c_e$. As for the growth of the average degree $\av{k(t_e,t)}$, we measure the average degree at time $t>t_e$ for all the nodes belonging to the class with entrance time $t_e$. In this way, we are defining a new set of classes only defined in terms of the entrance time $t_e$. The asymptotic behavior is found to be $\av{k(t_e,t)}\propto t^q$.

\subsubsection*{Model score}
We ran the model at different values of $R$ and $\rho$ for each one of the six sample strategies $s$ (see SI Section~\ref{sec:si_results} for details). For each dataset $d$ we select the configuration that best fit the data by looking at the score $S_d(\rho, R, s)$ that reads
\begin{equation}
    S^d(\rho, R, s) = \sum_{i=1}^8{\frac{|o^d_i - \tilde{o_i}(\rho, R, s)|}{\sigma^d_i}},
    \label{eq:score}
\end{equation}
where $o^d_i$ and $\sigma^d_i$ are the value and uncertainty on the $i$-th observable of the empirical dataset and $\tilde{o_i}(\rho, R, s)$ is the value of the same observable measured in the simulations with configuration $(\rho, R, s)$. The eight selected observables are: 1) the exponent $\gamma$ leading the growth of the number of edges $E(t)\propto t^\gamma$, 2) the optimal $\beta$ measured in the  strengthening function $p(k)$, 3) the average clustering coefficient $c$, 4) the exponent leading the growth of the average degree per node class $\av{k(e,t)}\propto t^q$, 4-8) the fractions $OO$, $OC$, $NO$, $NC$ of events allocated toward old/new link insisting or not on a open/closed triangle (see SI Section~\ref{sec:si_results} for details).

\subsubsection*{APS subsampling}

In the APS dataset we transform each paper published by $n$ authors in a sequence of $E=n(n-1)$ events with all the possible links between all the ordered couples of co-authors.
We then sample $l$ links over the $E$ possible links for each paper to be inserted in the total sequence $\mathcal{S}$. 
The results reported in the main text refer to $l=1$, and the reader can refer to the SI Section~\ref{sec:si_APS_subsample} for results with different values of $l$ and different strategies of subsampling (number of sampled links proportional to $E$).

\subsubsection*{Events on new-old and open-closed edges}

We count, for each logarithmically spaced time interval, the number of events happening on edges that are either old (already activated in the past), new (being activated now) and that happen to close a triangle (closed) or not (open). These four categories are then: the old open ($OO$), the old closed ($OC$), new open ($NO$) and new closed ($NC$) that we define as the fraction of events falling in each category per time range in the asymptotic limit of the system evolution ---i.e., after 60\% of the events passed.

\bibliographystyle{unsrt}
\bibliography{bibliography}

\end{document}



\maketitle

\section{Model}
\label{sec:si_model}

\subsection{Definition}
\label{sub:si_model_definition}

\begin{figure}
    \centering
    \includegraphics[width=1.0\textwidth]{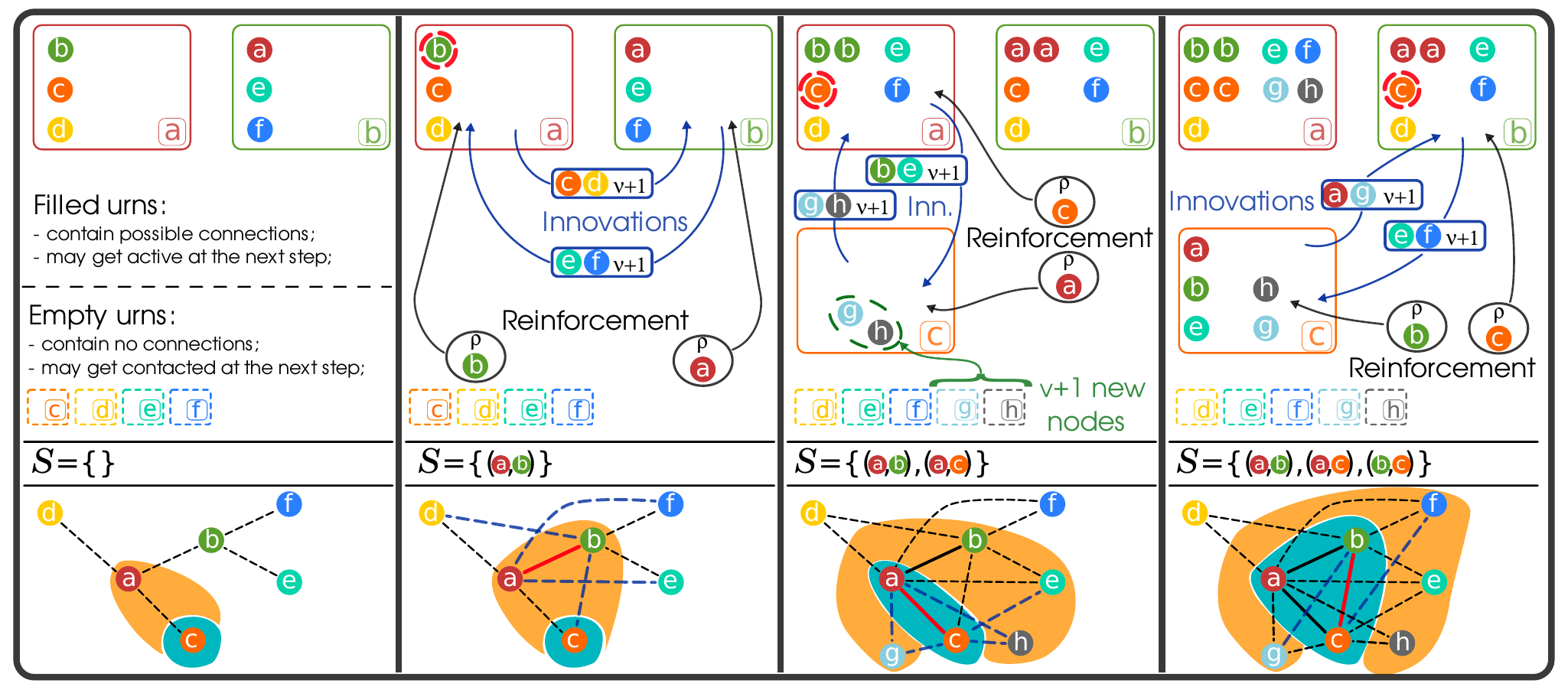}
    \caption{
        \label{fig:si_model} Three evolution steps of the Polya's urn model for a system with $\rho = \nu = 1$ with sampling strategy $s=$WSW. For each evolution step of the system (columns), we show the current state of the urns (top row), the sequence $\mathcal{S}$ of observed events (middle row) and the equivalent network evolution (bottom row). In the latter, we show the already active links (solid lines), links in the adjacent possible (dashed lines), currently active links (red lines) and connections entering into the adjacent possible (blue dashed lines). The shadowed areas correspond to the actual (cyan area) and the adjacent possible (orange area) of node $c$. {\bf First column}: $\mathcal{S} = \left\{\right\}$ is initially empty and we have only the $a$ and $b$ urns containing different IDs. Empty urns ($c-f$) are represented as dashed boxes. {\bf Second column}: we select the active urn proportionally to its size (urn $a$ in this case) and draw a ball with ID $b$ from it (circled with the red dashed line). We append the $(a,b)$ event to the sequence $\mathcal{S}$ (middle row) and evaluate the reinforcement-novelties steps of the model since the $e_{ab}$ link is new: we put $\rho$ copies of $a$ into urn $b$ and vice versa (reinforcement) and we draw $\nu+1$ distinct balls from $a$ following the WSW strategy (balls $c$ and $d$) that will be copied into $b$ (we do the same for $b$ that sends $e$ and $f$ as novelties to $a$). The effect on the growing network is the activation of the $e_{ab}$ edge and the promotion of the $e_{ae},\,e_{af},\,e_{bc}$ and $e_{bd}$ links to the adjacent possible (bottom row). Notice that the adjacent possible of $c$ changed without the need for $c$ to participate in a social interaction. {\bf Third column}: in the next event $a$ draws a copy of $c$ (top). Since $c$ is an empty urn, it creates $\nu+1$ novel IDs ($g$ and $h$) and gains a copy of them. We add $(a,c)$ to the sequence $\mathcal{S}$ (middle) and we perform the reinforcement/novelties exchange between $a$ and $c$. The network gains two new nodes ($g$ and $h$), activates a new edge ($e_{ac}$) and inserts new links in the adjacent possible. The actual space of $c$ acquires $a$ while her adjacent possible gains $e,\,g$, and $h$. {\bf Fourth column}: we extract $c$ from urn $b$ thus closing the $abc$ triangle. The link between $b$ and $c$ is again a new link so that we perform both the reinforcement and the novelties exchange steps. We add the $(b,c)$ event to $\mathcal{S}$ and add the new link in the network. Note that the $a-b-c$ triangle has been closed because $a$ recommended $b$ as a contact to $c$.}
\end{figure}

\begin{figure}
    \centering
    \includegraphics[width=0.75\textwidth]{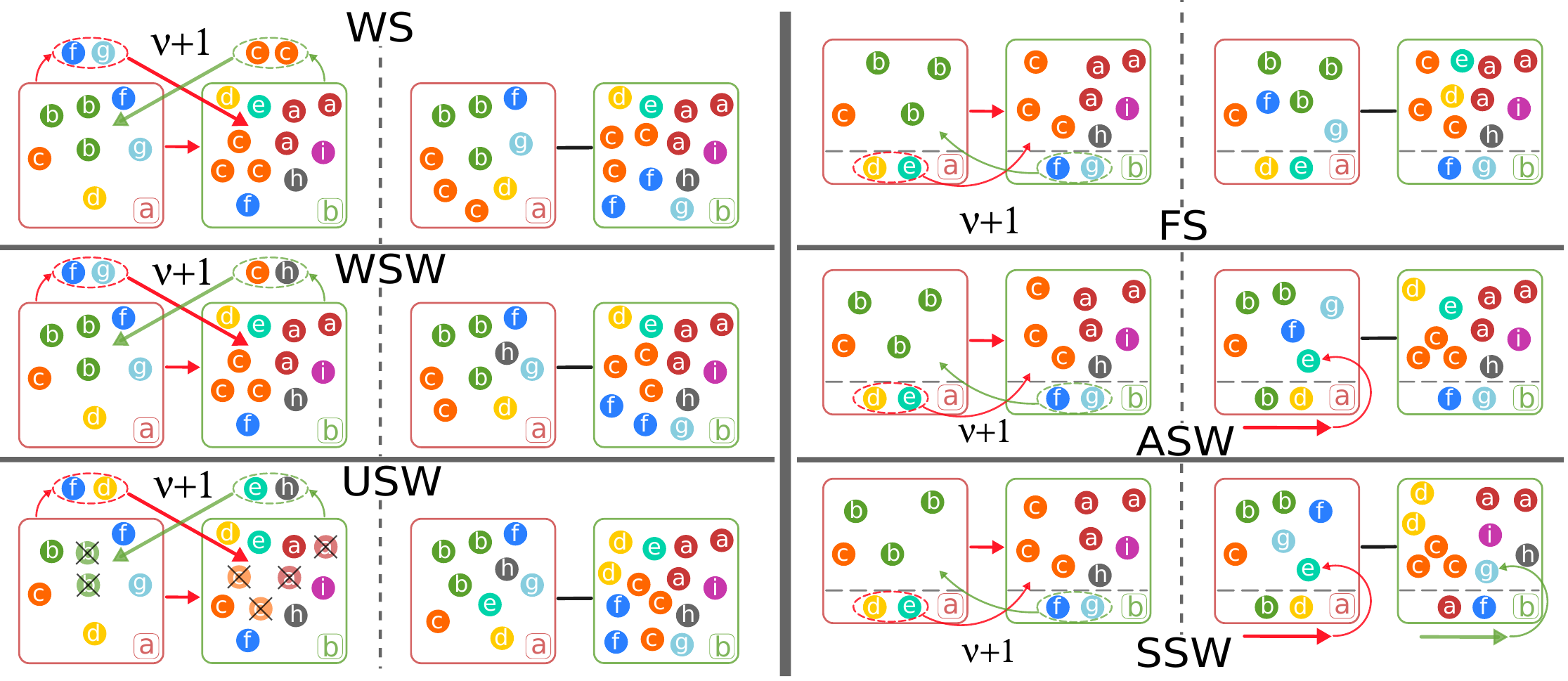}
    \caption{
        \label{fig:si_strategies} The six sample strategies used in the work in the $\nu=1$ case. For each strategy we show the status of two interacting urns $a$ and $b$ (where $a$ is actively contacting $b$) before and after interacting for the firs time. The balls being exchanged are highlighted within the dashed lines. Note that we do not show the reinforcement of $\rho$ balls to improve readability.
        (Top left) The Weighted Sample (WS) strategy: each urn samples $\nu+1$ balls (without replacement) and send a copy of them to the other. In this situation it is possible for an urn to send two or more copies of the same IDs to the other urn, as a single ID can be drawn multiple times. For example, in this case $b$ samples $c$ twice.
        (Center left) Weighted Sample with Withdrawal (WSW): each urn draws $\nu+1$ distinct IDs proportionally to their popularity in the urn (the more balls with same ID $k$, the more likely for an ID to be drawn). However, once a ball of ID $k$ has been drawn, all the copies with the same ID are withdrawn from the urn before the next ID is drawn. In this way we enforce the fact that exactly $\nu+1$ IDs will be exchanged.
        (Bottom left) Uniform Sample with Withdrawal (USW): as in the WSW case but all the IDs in an urn have the same probability to be drawn, i.e., we act as if there were only one ball for each ID by shadowing the redundant copies (gray crosses in the urns).
        (Top Right) Fixed Sons (FS) scheme: in this scheme and in the following ones, instead of extracting each time its memory buffer, each urn features a special set of balls that compose the memory buffer being exchanged. In this case the urns keep as their memory buffer the $\nu+1$ balls created when getting active for their first time (here $d$ and $e$ are the memory buffer of $a$ while $f$ and $g$ are the balls to exchange for $b$). The set of balls composing the memory buffer is static and never changes during the system evolution.
        (Middle right) Asymmetric Sliding Window (ASW): as outlined in the main paper each urn will pass the last $\nu+1$ contacted IDs. After the interaction has taken place only $a$ rotates her memory buffer by putting $b$ in front of it and downgrading $e$ as a regular ball in the urn.
        (Bottom right) Symmetric Sliding Window (SSW): the same as ASW but now both the urns rotates their memory buffer after the interaction.
    }
\end{figure}

We report here the extended definition of the model as reported in the main text and as shown in Fig.~\ref{fig:si_model}. The evolution model reads as follows:
\begin{enumerate}
    \item we start with $2$ urns, $a$ and $b$ that have one copy of the other agent in the urn ($a$ has one copy of $b$ and vice-versa); the urns also contains the $\nu+1$ distinct identities (IDs) representing the novelties each urn introduces in the system when she gets active. These $2(\nu+1)$ IDs are initially represented as empty urns; the events sequence $\mathcal{S}$ is initially empty (first column of Fig.~\ref{fig:si_model});
    \item for each time step we extract a ``calling'' urn $i$ with probability proportional to the urns size $n_i$, i.e., the number of balls contained in urn $i$, and then a ``called'' urn $j$ among the IDs in the calling urn. This ball $j$ is again extracted proportionally to the number of balls representing ID $j$ in urn $i$. This double extraction constitutes a single event $(i,j)$ that is appended to the total sequence $\mathcal{S}$ that now reads $\mathcal{S}=[(i,j)]$.
    \item irrespective of the $(i,j)$ interaction, we add $\rho$ balls with ID $i$ in the $j$'s urn and vice-versa;
    \item if it is the first time that $j$ gets called by another node it creates $\nu+1$ new agents (empty urns) into the system and a copy of each of them into the $j$'s urn; these $\nu+1$ IDs are, again, the novelties added to the system by the first activation of an empty urn ($j$ in this case) (third column of Fig.~\ref{fig:si_model});
    \item if it is the first time that $i$ and $j$ get in contact each of them samples $\nu+1$ balls accordingly to a strategy $s$ (see Fig.~\ref{fig:si_strategies} for details), and we add a copy of these balls into the other urn; depending on the strategy, it may happen that $i$ passes a copy of $j$ to $j$ itself (or the other way around). In that case we omit the copy of the $i$ ($j$) ball into the $i$'s ($j$'s) urn thus copying $\nu$ sons instead of $\nu+1$. This procedure avoid the possibility for a node to interact with itself (all the cases shown from the second to the last column of Fig.~\ref{fig:si_model});
\end{enumerate}

We tried six different sampling strategies $s$ in our work that are reported in Fig.~\ref{fig:si_strategies}.

Three of them feature a sampling of the urn content to compose the balls to be exchanged, whereas the others are designed to have a fixed or dynamical set of IDs to be exchanged along new connections.
Each strategy is designed to simulate different possible exploration mechanisms that agents adopt while they are probing their social space. Specifically, the Weighted Sample without replacement (WS) strategy requires an urn to sample $\nu+1$ balls from its content without replacement, thus sending each ID with a probability proportional to its prominence in the urn. This strategy leverages on the fact that the alters that have been contacted the most (i.e., the most represented in terms of balls given the numerous reinforcements) are the one more likely to be suggested as new contacts to the interacting urns seen for the first time. Moreover, an urn $i$ can withdraw an ID more than once so that the effective number of novelties sent to the interacting urn $j$ is less than $\nu+1$.
To relax the fact that the most reinforced alter is likely the only one proposed as a novelty to the interacting urns, we also implemented the Weighted Sample with Withdrawal (WSW) strategy, where after each one of the $\nu+1$ extractions we withdraw from the urn all the balls of the extracted ID, thus excluding it from the following extractions. In this way we enforce the selection of $\nu+1$ different IDs. This strategy preserves the weighted extraction (as every ID is drawn proportionally to the number of balls representing it) but enforces a fixed number of IDs to be exchanged between two urns. This strategy replicates the fact that an agent $i$ may recommend to another urn $j$ $\nu+1$ IDs selected proportionally to the number of times they have been in contact with $i$. The WSW strategy has been found to be optimal in the TMN dataset, where this mechanism may reflect the fact that two users interacting for the first time may exchange, on average, $\nu+1$ different IDs. These ID may be chosen by $j$ by browsing the $i$'s wall of posts, where the probability for a new account $k$ to be chosen by $j$ is proportional to the number of communication events that $k$ had with $i$ in the past, i.e., to the number of times $k$ appears in the wall of $i$.
To limit further the propensity of the most popular IDs to be chosen in the novelties sampling process we further introduced the Uniform Sampling with Withdrawal (USW) strategy that reads the same as the WSW one but where IDs are extracted uniformly, i.e., irrespective of their abundance in the urn. In this way, an element $k$ recently introduced into urn $i$ has the same probability to get extracted as the most reinforced ones.

The other three sampling strategies feature a deterministic set of balls to be exchanged between interacting urns during they first contact.
The basic strategy is the Fixed Sons scheme (FS), where each urn keeps the $\nu+1$ IDs created at its entrance in the system as a fixed memory buffer --- that is, her "sons". These will be the balls that will be exchanged by an urn $i$ with an urn $j$ getting in contact for the first time. This is a rigid scheme where the recommended/exchanged novelties of an urn do not evolve in time and do not get affected by the evolution of the system.
Again, to relax the assumption on the static nature of the memory buffer we introduce the Asymmetric Sliding Window strategy (ASW). In the latter we still have a set of balls that will be exchanged by the urn $i$ during the first contact with an urn $j$ but this set evolves in time. This group of balls is set to be the $\nu+1$ balls created by the urn $i$ at her entrance in the system and it is later updated during every new contact actively engaged by $i$ --- i.e., every contact where $i$ engages a communication toward $j$ --- by putting $j$ in front of the memory buffer and removing from it the oldest component of the set, that is, the $\nu+1$-th ball. This scheme promotes the diffusion of the recently contacted IDs rather than the mostly contacted ones, so that it enhances the probability for an ID entering the system at a later time to spread in different urns. The asymmetric nature of the updating rule (the "passive" element $j$ of a communication $i \to j$ does not rotate its memory buffer) reflect the fact that an individual $j$ must reciprocate an $i\to j$ interaction (i.e., we have to observe at least one reciprocal event $j \to i$) before accepting $i$ as a candidate to be shared with fresh contacts. This scheme is found to better fit the MPN dataset. This is reasonable as we expect people to share their last connections, that may in general differ with respect to their most contacted ones.

\subsection{Analytical results}
Given the definition of the model we can analytically tackle the analysis of the asymptotic behavior of the system evolution. To this end, we focus on two observables of the system, i.e. the number of distinct IDs $D(t)$ in the events sequence $\mathcal{S}$ and the number of edges (multiplied by two) $E(t) = \sum_{i=1}^{D(t)}{k_i(t)}$, where $k_i$ is the number of distinct IDs that have been in contact with node $i$ up to the evolution step $t$ (i.e. the cumulative degree of node $i$ at time $t$). Here we solve the problem with the $s=$FS strategy but, as we will show later, these results are quite robust with respect to the strategy change.

To write the time dependence of these two observables we must introduce some more quantities so as to correctly take into account the different contribution to the system evolution. In particular we define $F(t) = \sum_{i=1}^{D(t)}{k_{f_i}(t)}$, i.e. the sum for each node $i$ of the degree of its ``father'' (i.e. the urn that generated the $i$-th ID) $f_i$ at time $t$. As suggested from numerical simulations, this sum grows with the same time-dependence of $A(t)$ but with a different multiplying constant so that $F(t)=fA(t)$.
We also define $\tilde p$ to be the probability for an urn to be connected to her generator, i.e. for urn $j$ to be connected with the urn $i$ that created $j$ when entering the system. Prompted by numerical simulations we will assume this probability to be constant in time and node-independent.
Finally, we define $N(t)$ as the total number of balls in the system, i.e. $N(t) =\sum_{i=1}^{D(t)}{n_i(t)}$, where $n_i(t)$ is the number of balls in the $i$-th urn. Given the model definition we have:
\begin{equation}
    N(t) = N_0 + 2\rho t + (\nu+1) D(t) + (\nu+1) A(t),
    \label{eq:Nt}
\end{equation}
where $t$ is the number of evolution steps and $N_0 = 2 + 2(\nu+1)$ is the initial number of balls in the system.

Given these definitions, we can write the master equation governing the evolution of the $D(t)$ term that reads:
\begin{equation}
    \frac{d D(t)}{dt} = \frac{ (\nu - \tilde p) D(t) + (\nu+1-f) A(t)}{N(t)},
    \label{eq:me_Dt}
\end{equation}
where we took into account that the number of balls in the system that still did not appear in the sequence $\mathcal{S}$ is the number of created IDs $(\nu+1)D(t)$ plus the number of exchanged (and thus duplicated) novelties $(\nu+1)A(t)$ minus the copies of the IDs already present in the sequence $\mathcal{S}$. These are the $D(t)$ copies of the IDs extracted in $\mathcal{S}$ and the balls missing because being exchanged between a "son" and a "father" $\tilde p D(t)$ plus the $F(t)$ copies of each ID $i$ in $\mathcal{S}$ around the system spread by the contacts cumulated by the urn that generated $i$.

In the same way we can evaluate the master equation governing the evolution of the degree $k_i(t,t_i)$ of node $i$, i.e. the degree at time $t$ of an urn whose appearance time is $t_i$ (i.e. the event-time of the first appearance of urn $i$ in the sequence $\mathcal{S}$).

To write the equation governing the $k_i(t,t_i)$ evolution we have to account for the different contributions to the possibility for the urn $i$ to contact (or get contacted by) a new ID in the network.
In particular this probability is proportional to the number of ``new'' balls in the $i$-th urn $(\nu+1)(1+k_i)$, i.e. the sons created at the appearance time and the copies of the sons got from the $k_i(t,t_i)$ established ties. In addition we have to sum the $k_{f_i}(t,t_{f_i})$ copies of the $i$-th ID that its father spread around the network.

We then have to subtract the over-counted balls, in particular the $k_i(t,t_i)$ copies of $i$ and of its sons that are no more ``new'' as for each established tie we burn a copy of an ID. We also have to account for the possibility for the node to have contacted its father $\tilde p$ that results in the loss of one ball. Gathering all the outlined terms we get:
\begin{equation}
    \frac{d k_i(t,t_i)}{dt} = \frac{(\nu + 1 - \tilde p) + (\nu + f) k_i(t,t_i)}{N(t)},
    \label{eq:me_kit}
\end{equation}
being $t_i$ the entrance time of the $i$-th urn in the system. Considering the boundary condition $k_i(t=t_i, t_i)=1$ and approximating $N(t)\simeq 2\rho t$ the solution of Eq. (\ref{eq:me_kit}) reads:
\begin{equation}
    k_i(t,t_i) = -\frac{(\nu+1-\tilde p)}{(\nu+f)}
                + \mathcal{C} t^{(\nu+f)/(2\rho)},
    \label{eq:sol_kit_long}
\end{equation}
where $\mathcal{C} = [1 + (\nu+1-\tilde p) / (\nu+f)]/t_i^{(\nu+f)/(2\rho)}$, so that:
\begin{equation}
    \begin{split}
        k_i(t,t_i) = -\frac{\nu+1-\tilde p}{\nu+f}+
        \left( 1 + \frac{\nu+1-\tilde p}{\nu+f}\right)
             \left(\frac{t}{t_i}\right)^{(\nu+f)/(2\rho)} =\\
    = -\mathcal{Q} +
    \left( 1+\mathcal{Q} \right) \left(\frac{t}{t_i}\right)^{(\nu+f)/(2\rho)},
    \label{eq:sol_kit}
    \end{split}
\end{equation}
where we set $\mathcal{Q}=(\nu+1-\tilde p)/(\nu+f)$. Now we can evaluate the $A(t)$ sum by substituting the just found functioning of $k_i(t,t_i)$:
\begin{equation}
    A(t) = \sum_{i=1}^{D(T)}{k_i(t,t_i)} \simeq
    \int_0^t{k(t,t') \frac{\partial D(t')}{\partial t'} dt'},
    \label{eq:eval_At_1}
\end{equation}
where we dropped the $i$ index from $k(t,t')$ and where we introduced the number of urns that entered the system at $t'$ as the differential of the number of urns $D(t)$. Prompted by numerical simulation we set $D(t)=gt^\gamma$, where $g$ is an
unknown constant and $\gamma$ the exponent leading the time evolution of the number of urns $D(t)$. By substituting Eq. (\ref{eq:sol_kit}) and the $D(t)$ form in Eq. (\ref{eq:eval_At_1}) we find:
\begin{equation}
    \begin{split}
    A(t) = \gamma g \int_0^t{k(t,t') t'^{\gamma-1} dt'} =
    \gamma g \int_0^t{\left[ -\mathcal{Q} + \left( 1+\mathcal{Q} \right)
    \left( \frac{t}{t'} \right)^{(\nu+f)/(2\rho)}\right] t'^{\gamma-1} dt'} =\\
    = g\left[ \frac{\gamma +  \mathcal{MQ}}{\gamma - \mathcal{M}} \right] D(t) = r(\gamma) D(t),
    \end{split}
    \label{eq:eval_At_2}
\end{equation}
where $\mathcal{M} = (\nu+f)/(2\rho)$. Eq. (\ref{eq:eval_At_2}) tells us that $A(t)$ evolves in time with the same exponent of $D(t)$ and the two are bound by a proportionality constant $r(\gamma)$ so that $A(t)=r(\gamma) D(t)$.
We can now solve the system by substituting Eq. (\ref{eq:eval_At_2}) in Eq. (\ref{eq:me_Dt}) getting:
\begin{equation}
    \frac{dD(t)}{dt} = \frac{(\nu-\tilde p)+(\nu+1-f)r(\gamma)}{2\rho t} D(t),
    \label{eq:sol_Dt}
\end{equation}
where, again, we approximated $N(t)\simeq 2\rho t$ as we expect $\gamma < 1$. By substituting $D(t)=qt^\gamma$ in Eq. (\ref{eq:sol_Dt}) we get a second-order equation whose positive solution gives us the predicted value of $\gamma$ that reads:
\begin{equation}
    \gamma = \frac{\mathcal{B} +
    \sqrt{\mathcal{B}^2 + 8\rho\mathcal{M} [\tilde p - \nu + (\nu+1-f)\mathcal{Q}]}}{4\rho}
    \xrightarrow[]{\rho,\nu\to\infty} \frac{3}{2}{\nu\over\rho} =
    {3\over2}\mathcal{R}^{-1},
    \label{eq:sol_gamma_1}
\end{equation}
where $\mathcal{B} = (2\nu + 2\rho\mathcal{M} + 1 - f - \tilde p)$ and where we introduced the ratio $\mathcal{R}=\rho/\nu$. Note that the solution of Eq.(~\ref{eq:sol_gamma_1}) holds in the $\mathcal{R}>3/2$ region. For $\mathcal{R}\le3/2$ we cannot approximate $N(T)\simeq 2\rho t$ as the $D(t)$ and the $A(t)$ terms are now comparable to the linear in time term $2\rho t$. In this case one should solve the set of coupled equations:
\begin{equation}
    \begin{cases}
        {dk(t,t')\over dt} = \frac{\nu+1-\tilde p + (\nu+f)k_i(t,t')}
            {2\rho t + (\nu+1)[A(t)+D(t)]} \\
        A(t) = \int_0^t{k(t,t') {dD(t')\over dt'} dt'} \\
        {dD(t)\over dt} = \frac{(\nu-\tilde p)D(t) + (\nu+1-f)A(t)}
            {2\rho t + (\nu+1)[A(t)+D(t)]}.
    \end{cases}
    \label{eq:ratio_ge_1.5}
\end{equation}

We can now go back to Eq. (\ref{eq:eval_At_2}) and substitute Eq.
(\ref{eq:sol_gamma_1}) therein to get the proportionality constant $r(\gamma)$
between $D(t)$ and $A(t)$ in the $\nu,\rho\to\infty$ limit:
\begin{equation}
    r(\gamma) \to { {3\over2}\mathcal{R}^{-1} + {1\over2}\mathcal{R}^{-1}
                \over {3\over2}\mathcal{R}^{-1} - {1\over2}\mathcal{R}^{-1}} =
                2.
    \label{eq:const_r}
\end{equation}
Eq. (\ref{eq:const_r}) tells us that the proportionality constant does not depend (in the $\rho\gg1$ limit) on the ratio $\mathcal{R}$.

\section{Data}
\label{sec:si_data}

\subsection{\textit{A}merican \textit{P}hysical \textit{S}ociety}
\label{sub:si_aps}

The APS dataset logs the co-authorship networks found in the Journals of the American Physical Society covering the period between January 1970 to December 2006 and contains 301,236 papers written by 184,583 authors that are connected by 995,904 edges~\cite{radicchi_scientific_2009}.

In this dataset each interaction represent the authors of a paper that is being published. Since we cannot give a directionality to the data we transform the sequence of papers to a sequence of bi-grams: for a paper with authors ${a,b,c}$ we insert in the sequence of data all the possible permutation of two authors, i.e. ${(a,b), (a,c), (b,c), (b,a), (c,a), (c,b)}$. Since our time resolution is limited to the day in which a specific issue of a given journal was published (and we cannot put a meaningful time order within a single journal issue) we grouped all the journals issues by their year, month and decade (i.e., we group together all the issues published between the 1st and the 10th, the 11th and the 20th, the 21st and the last day of month for each month in each year).

When analyzing this dataset we define the user's activity $a_i$ as the number of times he appears as the initiator of an interaction. For example, an author $i$ that publish two  papers, the first with 3 co-authors and the second with a single co-author, has activity $a_i = 4$.

We do not include large collaborations in our analysis (papers with more than ten authors). Details on the applied procedure to get the data and perform name disambiguation can be found in~\cite{radicchi_scientific_2009}.

\subsection{\textit{T}witter \textit{M}ention \textit{N}etwork}
\label{sub:si_tmn}

The dataset of \textit{Twitter} is composed by $273$ daily files covering the period between January the $1^{\rm st}$ to September the $30^{\rm th} 2008$ containing the \emph{fire-hose} of the platform, i.e., all the $16,329,466$ citations done by all the $536,210$ users in the given period. The nodes in the network are connected via $2,620,764$ edges.

In this work:
\begin{itemize}
\item we consider all the citations performed by all the users on the platform in the selected period, without discarding any of the collected event;
\item we define the activity $a_i$ of user $i$ as the number of mentions performed by $i$, i.e. the number of events actually engaged by the node $i$.
\end{itemize} 

\subsection{\textit{M}obile \textit{P}hone \textit{N}etwork}

The dataset of the \textit{M}obile \textit{P}hone \textit{N}etwork (\textit{MPN}) is composed by a single file containing the $1,949,624,446$ time ordered events with $1$ second resolution covering the period between January and July of $2008$ for $6,779,063$ users of a single operator with $20\%$ market share in an European country.

The dataset contains all the events involving users of the company, then we also have the calls from non-company users to company users and vice-versa.
In total, we have $33,160,589$ nodes (of which $6,779,063$ are company users) connected through $92,784,825$ edges.

While we consider all the time-ordered events for this dataset, we limit the measures of all the observables to the nodes being users of the company.
Also, when counting the clustering coefficient, we limit the measure of the possible triangles to the edges where both ends belong to the company (as we cannot observe links between non-company users).

In this dataset the activity of an user is defined as the number of calls $a_i$ actually engaged by the node $i$.

\subsection{Synthetic data}

The model defined in the main text and below in Section~\ref{sec:si_model} naturally outputs a sequence of events of the type $({\rm ID}_{\rm from}, {\rm ID}_{\rm to})$ that can be naturally interpreted as a sequence of events.
Here the activity of an ID $i$ is the number of times node $i$ appears as the ID initiating the interaction (i.e. being the ${\rm ID}_{\rm from}$).

\section{Results}
\label{sec:si_results}

\subsection{Model properties}
\label{sub:si_model_props}

We report here the results concerning the model properties. In particular we focus on the dependence of different observables on the choice of the three model parameters ($\rho,\,R=\rho/\nu,\,s$).

\begin{figure}
	\centering
    \includegraphics[width=.75\textwidth]{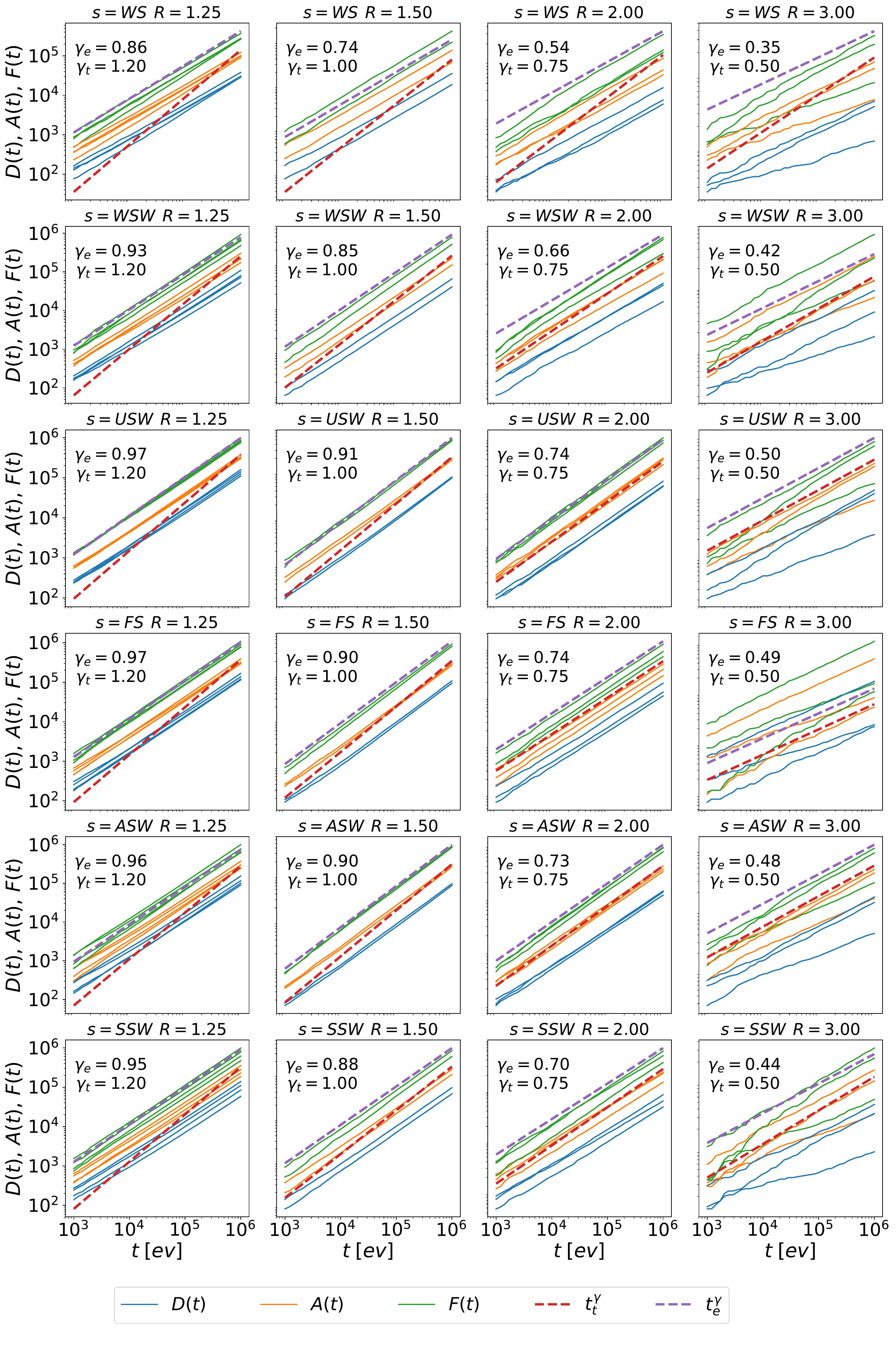}
    \caption{
        \label{fig:si_model_growth}
        The $D(t)$ (blue lines), $A(t)$ (orange lines) and $F(t)$ (green lines) as measured from synthetic data for different sampling strategies $s$ (one per row) and diverse values of ratio $R$ (one for each column. Within each plot we show diverse curves for different values of $\rho$ at fixed strategy $s$ and ratio $R$ together with the theoretical exponent $\gamma$ of Eq.~\ref{eq:sol_gamma_1} (red dashed lines) and the empirical one (purple dashed lines) whose values are reported as text in the subplots.
    }
\end{figure}

\begin{figure}
    \centering
    \includegraphics[width=.85\textwidth]{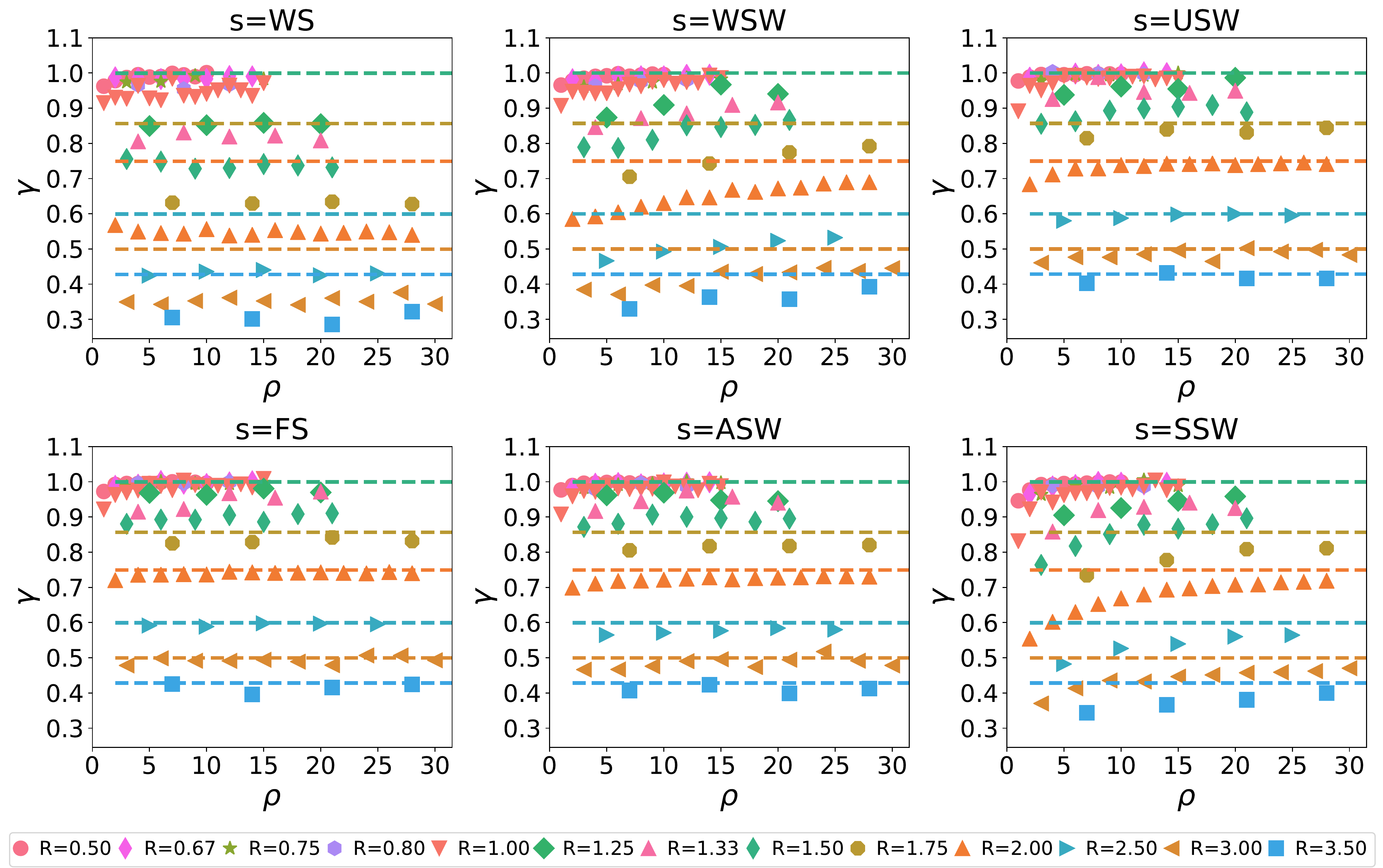}
    \caption{
        \label{fig:si_model_gamma}
        The $\gamma$ exponent leading the growth of $D(t)$ as found for different strategies $s$ (one for each subplot) and diverse values of ratio $R$ (colored markers). Within each plot, for the $R\ge 1.5$ curves, we show diverse curves for different values of $\rho$ at fixed strategy $s$ and ratio $R$ together with the asymptotic theoretical exponent $\gamma = 3/2/R$ of Eq.~\ref{eq:sol_gamma_1} (dashed lines, same color of the same ratio's markers).
    }
\end{figure}

\begin{figure}
    \centering
    \includegraphics[width=.75\textwidth]{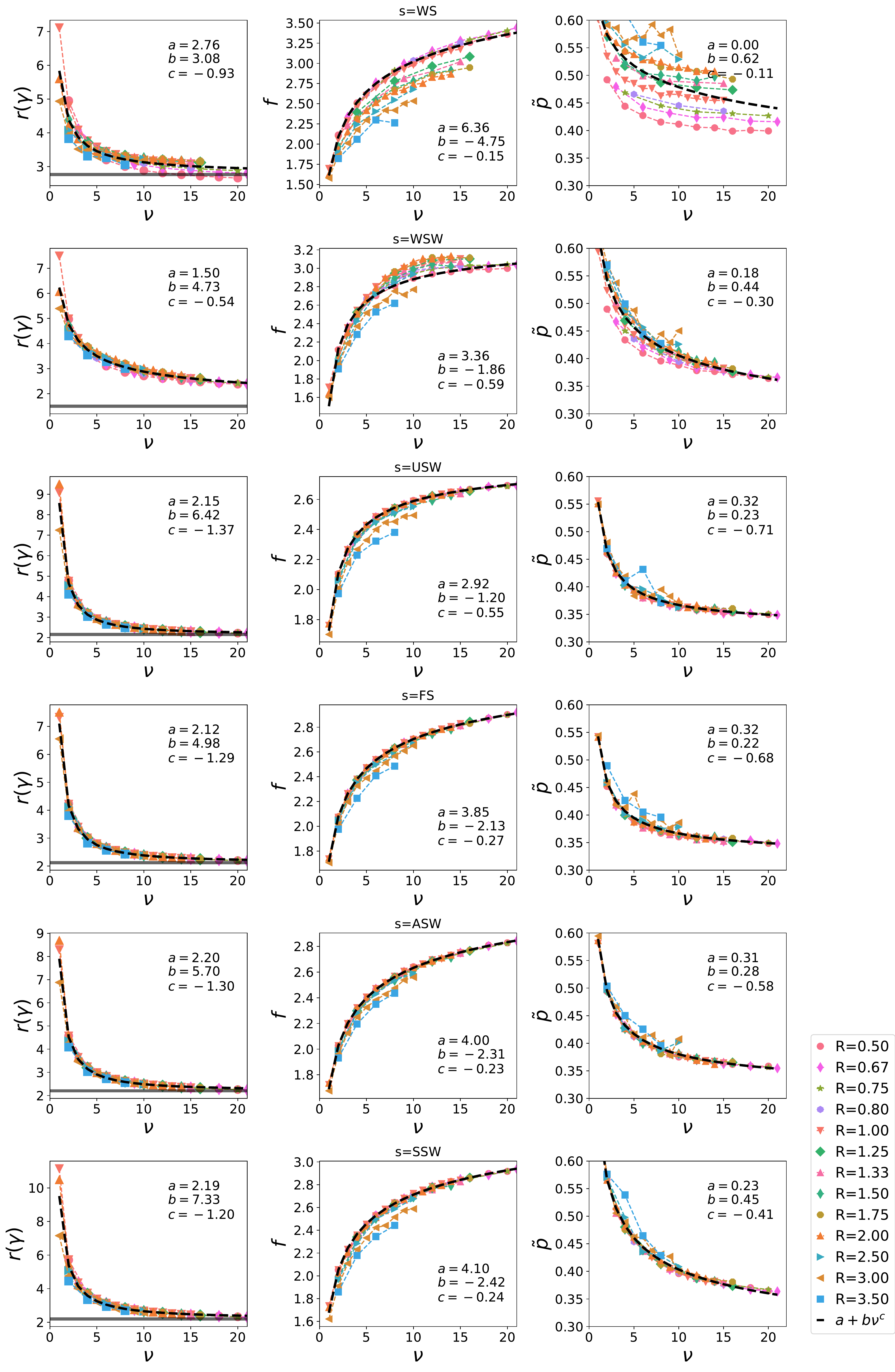}
    \caption{
        \label{fig:si_model_constants}
        (First column) The $r(\gamma)$ constant linking the $A(t)$ with the $D(t)$, (second column) the $f$ constant fixing the $F(t)=fA(t)$, and, (third column) the probability $\tilde{p}$ for an urn to be connected to her generator (or the "father" urn) for different strategies (rows from top to bottom) as measured from synthetic data for different ratio $R$ (symbols). The data are reported as a function of $\nu$ as all the curve collapse to a single behavior $y = a+b\nu^c$ (black lines, parameters reported as text in each subplot). For the $r(\gamma)$ we also show the predicted theoretical asymptotic value $r(\gamma) = 2$ for $\nu,\rho\to\infty$ (black solid line).
    }
\end{figure}

First, in Fig.~\ref{fig:si_model_growth} we show that the assumptions made in the analytical calculations are correct, as all the relevant observables ($A(t)$, $D(t)$ and $F(t)$) all grows at the same pace $t^\gamma$ and they differ only by a proportionality constant.
We then first check in Fig.~\ref{fig:si_model_gamma} that the growing exponent $\gamma$ reads as in Eq.~\ref{eq:sol_gamma_1}, finding a good agreement between the empirical case and the theoretical predictions for $R\ge 1.5$.
In Fig.~\ref{fig:si_model_constants} we then show the behavior of the proportionality constants involved in the analytical solutions. Specifically, we show the constant $r(\gamma)$ setting the proportionality constant between $A(t)$ and $D(t)$ as found in Eq.~\ref{eq:eval_At_2}, and the constant $f$ linking $F(t) = fA(t)$. We also show the behavior of the $\tilde p$ constant that measures the probability for an urn to be connected to the urn that introduced her in the system (i.e., her "father").
Surprisingly, we find that all the constants measured for system featuring at different ratios $R$ but with a common strategy $s$ all behave similarly when plotted against $\nu$, i.e., $y=a+b\nu^c$. In particular the $r(\gamma)$ decreases as $\nu$ increases, meaning that at fixed ratio the average degree $\av{k}$ of the system is lowering (since $\av{k}\propto A(t)/D(t)$). On the other hand the number of links $F(t)$ emanating from the active urns that introduced IDs in the system (the urns that are father to some ID) is increasingly higher as $\nu$ increases. Lastly, the probability $\tilde{p}$ for an urn to be connected with her father decreases with $\nu$, as the number of possible connections that an urn being generated may activate increases with $\nu$ thus decreasing the probability to contact her father (the urn that introduced her in the system).

\begin{figure}
	\centering
    \includegraphics[width=.75\textwidth]{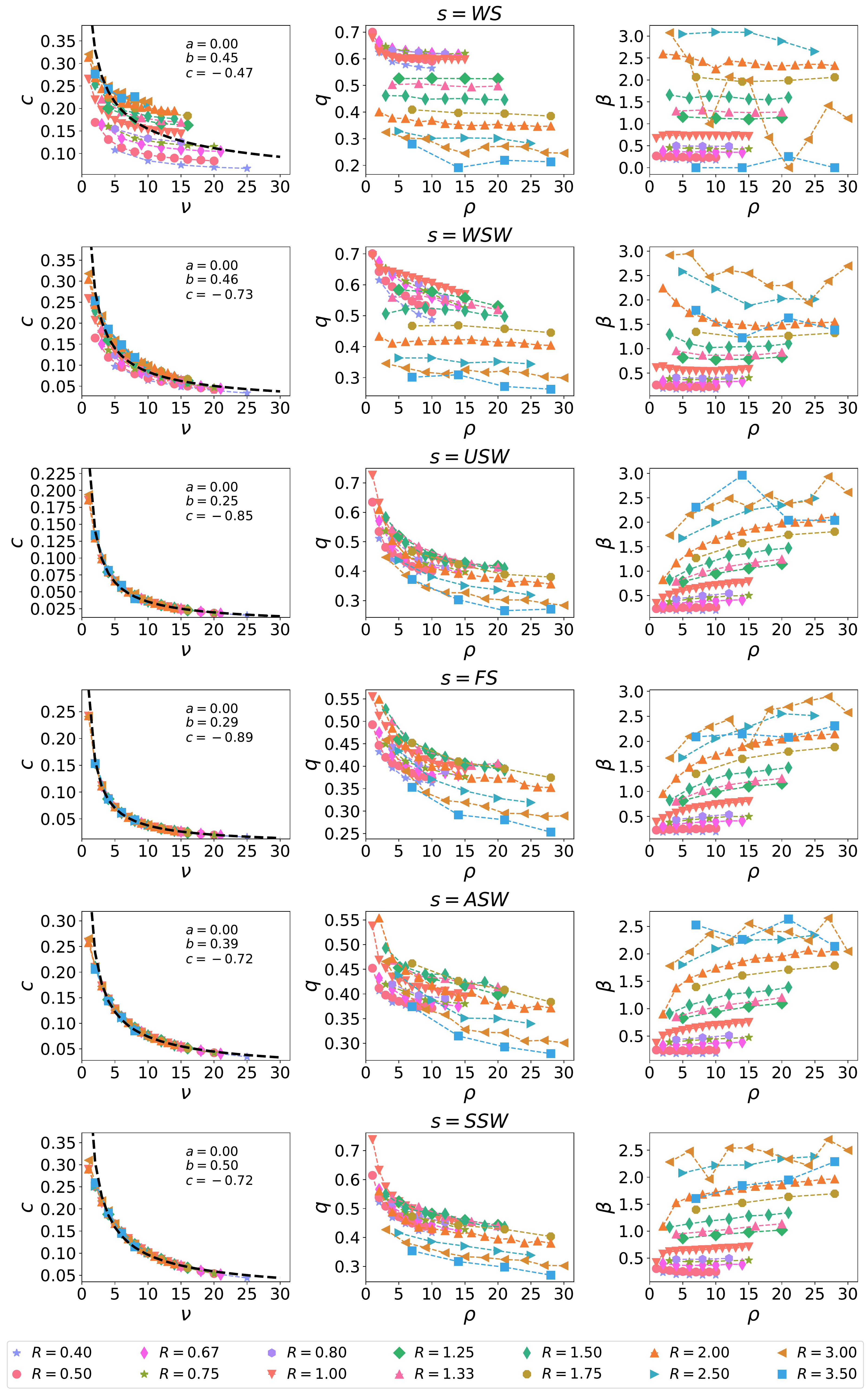}
    \caption{
        \label{fig:si_model_clustKatBeta}
        (First column) The average local clustering coefficient $c$, (second column) the $q$ exponent leading the $\av{k(t)}\propto t^q$, and, (third column) the strengthening exponent $\beta$ for different strategies (rows from top to bottom) as measured from synthetic data for different ratio $R$ (symbols). The data are reported as a function of $\rho$ except for the first column where we plot it against $\nu$ as all the curve collapse to a single behavior $y = a+b\nu^c$ (black lines, parameters reported as text in each subplot).
    }
\end{figure}

\begin{figure}
	\centering
    \includegraphics[width=.75\textwidth]{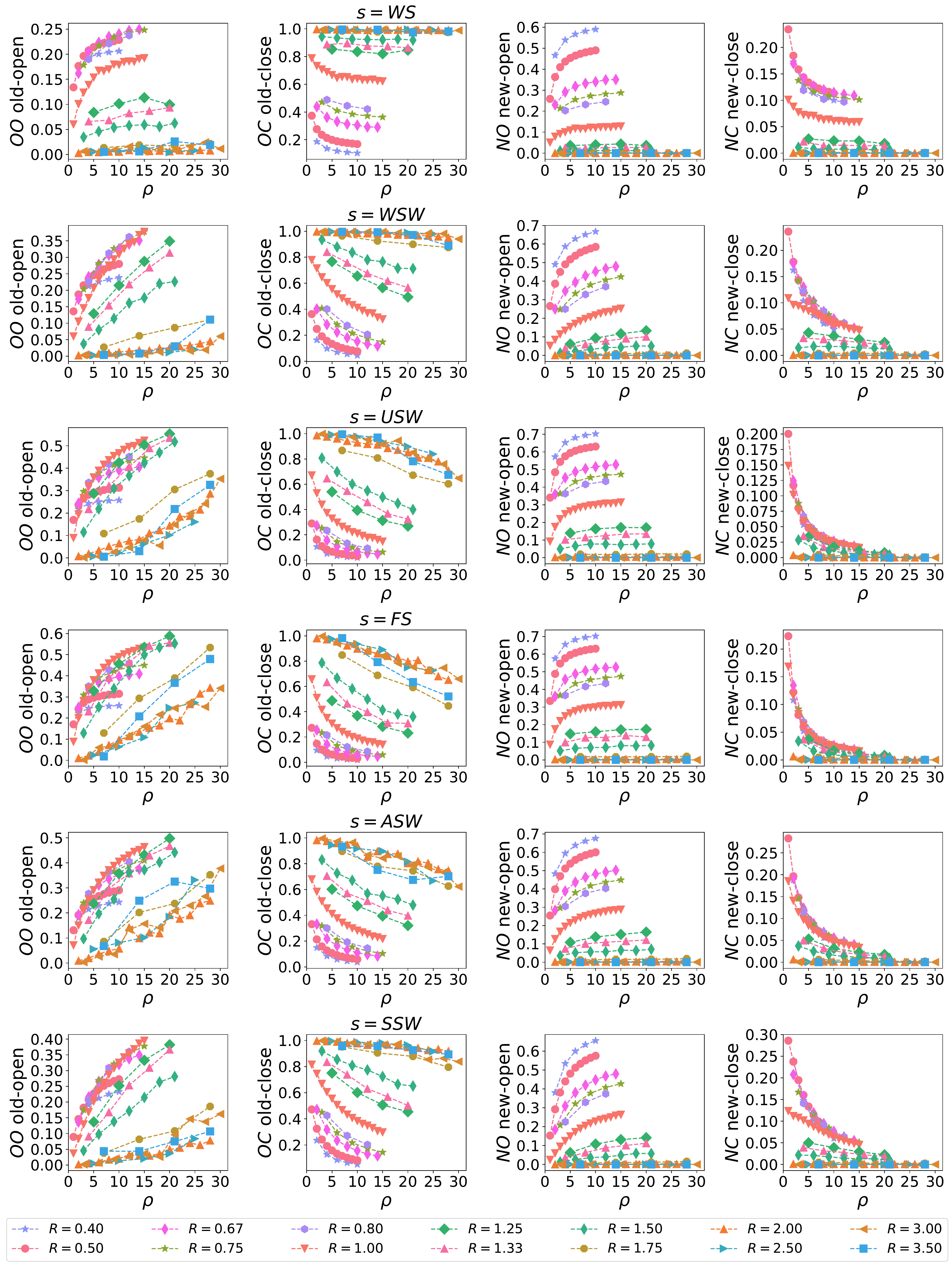}
    \caption{
        \label{fig:si_model_closure}
        For each column we show the empirical fraction (in the long time limit) of events insisting on a old open edge $OO$, old closed link ($OC$), new open edge ($NO$) and new close $NC$ edge (symbols) for different strategies (rows from top to bottom) as measured from synthetic data for different ratio $R$ (symbols) as a function of $\rho$.
    }
\end{figure}

The same functioning is found in the average clustering coefficient as shown in the first column of Fig.~\ref{fig:si_model_clustKatBeta}. In the same figure we also show the dependence on the model parameters of the average degree growth exponent $q$ and the strengthening exponent $\beta$.
Finally, in Fig.~\ref{fig:si_model_closure} we show the behavior of the fraction of events happening along links that are either old (already activated in the past), new (being activated during the event) and that insist on closed or open triangles in the long time limit.

\begin{figure}
    \centering
    \includegraphics[width=.8\textwidth]{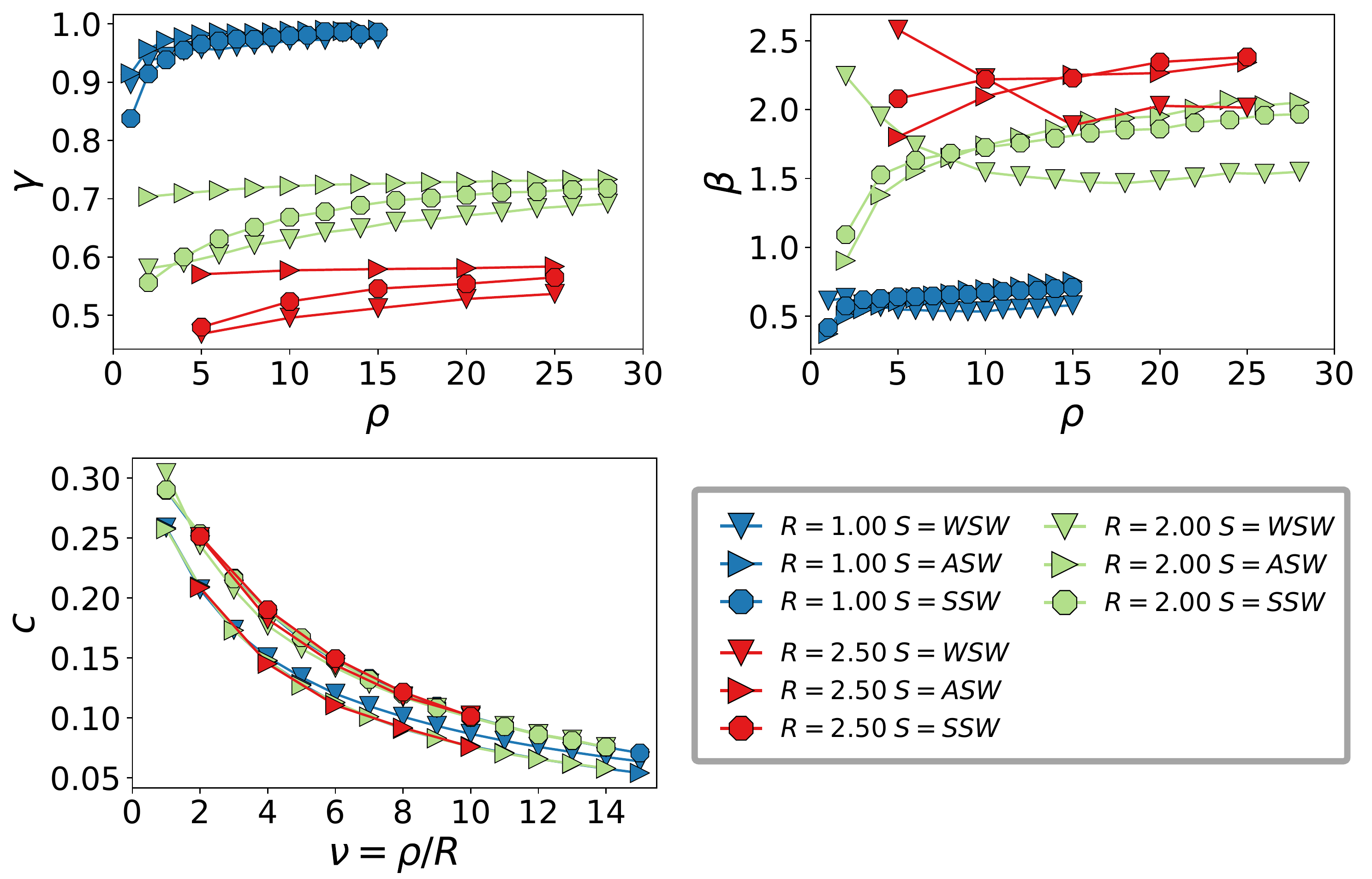}
    \caption{
        \label{fig:si_paramsInfluence} The influence of the three model parameters on a selected set of observables: (A) the $\gamma$  exponent leading the $A(t)\propto t^\gamma$ growth, (B) the strengthening exponent $\beta$, and, (C) the average local clustering coefficient $c$. In each panel we show the dependence on $\rho$ (in C we plot against $\nu$ as all the curves collapse on a single behavior) for three different ratio values $R=[1, 2, 2.5]$ (red, orange and cyan symbols) as well as for three different strategies $s = [$WSW, ASW, SSW$]$ (triangles, right triangles and circles).
}
\end{figure}

To sum up, in Fig.~\ref{fig:si_paramsInfluence} we compare in a single place how the model parameters affect three main observables of the system, nemly the $\gamma$ exponent leading the $A(t)\propto t^\gamma$ growth of the nodes and edges of the system in time, the strengthening exponent $\beta$, and the average clustering coefficient $c$. In Fig.~\ref{fig:si_paramsInfluence}A we show the values of the exponent $\gamma$ as a function of $R$, $\rho$ and the strategy $s$: $\gamma$ decreases as the ratio $R$ increases, whereas it weakly depends on both the reinforcement $\rho$ and the selected strategy $s$. Indeed, a larger $R$ corresponds to a stronger reinforcement of the links entering the system in its early stage, thus inhibiting the creation of new edges.
The exponent $\gamma$ does not strongly depend on the strategy $s$. However, it appears to be systematically larger when $s=$ASW.
In Fig.~\ref{fig:si_paramsInfluence}B shows that the strengthening exponent $\beta$ ---setting the decrease rate of the probability $p(k)$ to acquire a new acquaintance--- increases with the ratio $R$ and with the reinforcement parameter $\rho$, while it weakly depends on the strategy $s$. Again, the higher reinforcement of existing links favours the strengthening of already established links while lowering the probability for an urn to select connections not yet explored.
Finally, in Fig.~\ref{fig:si_paramsInfluence}C we show that the average clustering coefficient $c$ weakly depends on both the strategy $s$ and the ratio $R$, while it strongly depends on $\rho$ (or $\nu$). Indeed, at fixed $R$, the higher the $\nu$, the higher the number of novelties that a node $i$ receives from and sends to a newly contacted node $j$: this results in an increased number of possible triangles in the network, lowering the fraction of actually closed ones.

To conclude this part, let us note that, despite the limited number of parameters, our model is flexible enough to produce a wide range of phenomenologies, from highly exploratory situations (high $\gamma$ and $\beta$ exponents) to more exploitative scenarios, with a reduced number of connections being explored.

\subsection{Node binning}
\label{sub:si_binning}

To average the dynamical observables over homogeneous groups of nodes we perform a hierarchical binning of the nodes.
First we divide nodes in $E$ classes $e={1,2,\ldots, E}$ depending on their entrance time $t_i$, being $t_i$ measured in intrinsic time, i.e., the number of events.
For each dataset we define $E$ logarithmically spaced bins between $t=1$ and $t=T+1$, where $T$ is the total number of events. We then assign each node to a class $e$ to be the index of the bins in which the time of entrance $t_i$ of node $i$ falls, i.e., $e | t_{e} <= t_i < t_{e+1}$.
This is the binning used to evaluate the average degree growth $\av{k_i(t)}_{i\in e}$ per class.
This gives us an idea about the time evolution of the average degree for the nodes that entered the system together but the class contains diverse degree values that may significantly differ. 
That is why, when we measure the strengthening probability $p(k)$, we perform an additional binning for the nodes within class $e$ by further diving them in $G_e$ groups depending on the final degree $k_i(T)$ of each node accordingly to a logarithmically spaced binning between the lowest $k^{\rm min}_e = \min(k_i)_{i\in e}$ and larger $k^{\rm max}_e = \max(k_i)_{i\in e}$ of class $e$. In this way each node is assigned to a class $(e,g)$, with $g = {1,2,\ldots,G_e}$.

\subsection{Strengthening function measure}
\label{sub:si_reinforcement_measure}

In the following, for clarity, we define $b$ as a unique identifier of the $(e,g)$ group of nodes determined by their entrance time and group of final degree.
To measure the link strengthening process of each system, we count all the communication events $a_b(k)$ engaged by every node $i$ of the $(e,g)$-th group while having degree $k_i = k$ ---$a_b(k)$ is the total number of events engaged by nodes of that class at degree $k$.
If a node $i$ of the $b$-th class engages an event leading to a degree increase $k_i = k\to k_i = k+1$, we increment the counter $n_b(k)$ by $1$, being $n_b(k)$ the total number of events that the nodes belonging to the $b$-th group with instant degree $k$ performed toward a new node. Conversely, when a node increases its degree because it passively received a new contact, the $n_b(k)$ counter is not incremented.

The best estimate of the probability for a new node to establish a new connection at degree $k$ then reads~\cite{Ubaldi_asymptotic_2016}:
\begin{equation}
    f_b(k) = {n_b(k) \over e_b(k)},
    \label{eq:pk_data}
\end{equation}
where $n_b(k)$ and $e_b(k)$ are the event counters as defined above.
We can also write an estimate of the uncertainty on $f_b(k)$ by assuming that there are no correlations between users and by checking that $1\ll n_b(k)\ll e_b(k)$.
In this case, the standard deviation $\sigma(f_b(k))$ on the estimate of $f_b(k)$ is:
\begin{equation}
    \sigma(f_b(k)) = \sigma_b(k) = \sqrt{f_b(k)(1 - f_b(k)) \over e_b(k)}.
    \label{eq:sigma_data}
\end{equation}

We then fit the $f_b(k)$ measured values with the proposed strengthening function $p_{b}(k, \beta)$:
\begin{equation}
    p_b(k, \beta) = \left( 1 + \frac{k}{c(b)} \right)^{-\beta},
    \label{eq:pn_chi2}
\end{equation}
where $c(b)$ is the strengthening constant of the $b$-th bin, $k$ is the cumulative degree and $\beta$ is the strengthening exponent. The fit is not straightforward as the $c$ and $\beta$ parameters are highly correlated.

The procedure is then to keep $\beta$ fixed, fit the curve varying the strengthening constants $c$, compute a squared residuals sum $\chi^2_b(\beta)$ and then try other $\beta$ values for all the groups of nodes in the system.
In particular, for each class $b$ and with a fixed $\beta$, we optimize the parameter $c(b)$, by minimizing the function $\chi^2_b(\beta)$:
\begin{equation}
    \chi^2_b(\beta) = \sum_{k = 1}^{K_b}{\frac{\left[ f_b(k) - p_b(k,\beta)
    \right]^2}{\sigma_b(k)^2}},
    \label{eq:chi2_single_curve}
\end{equation}
where the index $k$ runs over the $K_b$ points of the $b$-th bin's curve and $\sigma_b(k)$ is as defined in Eq. (\ref{eq:sigma_data}). By repeating this procedure for each value of $\beta\in[0, 5.0]$ we find, for each class $b$, a $\chi^2_b(\beta)$ curve.

To measure the $\beta_{\rm opt}$ parameter, we define the total mean square deviation $\chi^2(\beta)$ as:
\begin{equation}
    \label{eq:chi2_beta}
    \chi^2(\beta) = \sum_{b = 1}^{N_b}{\left[ \chi^2_b(\beta) \right]},
\end{equation}
where $N_b$ is the total number of curves, i.e. the number of node groups $b$.
We then define the optimal $\beta_{\rm opt}$ as $\beta$ value minimizing the function $\chi^2(\beta)$:
\begin{equation}
    \label{eq:beta_opt}
    \beta_{\rm opt} = \rm{min}_\beta(\chi^2(\beta)).
\end{equation}
The distribution of the strengthening constants $P(c_b)$ for each group of nodes $b$ is shown in Fig.~\ref{fig:results} of the main text. The value of $c_b$ is determined by looking at the $c_b$ value found when fitting the $p(k)$ strengthening function for the group of nodes $b$ fixing $\beta=\beta_{opt}$.

\begin{figure}
    \centering
    \includegraphics[width=.9\textwidth]{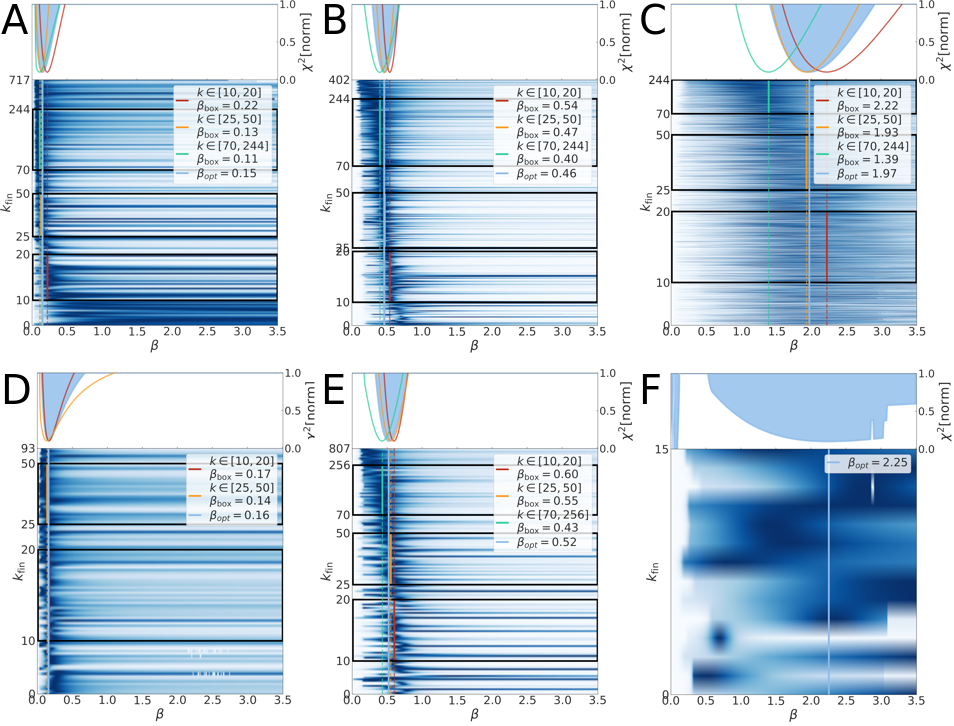}
    \caption{
        \label{fig:si_HM_data}
        The heat-map-like value of $\chi^2_{\rm opt}(b)/\chi^2_b(\beta)$  (bottom plots). We plot the exponent $\beta$ on the $x$-axes and the different classes of nodes $b$ sorted by their final degree on the $y$-axes. The color-map is proportional to $\chi^2_{\rm opt}(b)/\chi^2_b(\beta)$ representing the goodness of fit: the darker, the higher. The cyan vertical line is the value of $\beta_{opt}$, while the other vertical lines represent the same quantity evaluated in the three black boxes corresponding to different final degree intervals.
        (Top plots) The curve $\chi^2(\beta)$ as defined in Eq. (\ref{eq:chi2_beta}) (up-filled curve) and the same quantity for the three final degree intervals. For (A) APS $\beta_{opt} = 0.15$, (B) TMN $\beta_{opt} = 0.46$, (C) MPN $\beta_{opt} = 1.97$. We then show the three best fitting models for APS (D) $\rho=6,\,R=0.4,\,s=$SSW giving $\beta_{opt} = 0.16$, the model fitting the TMN (E) $\rho=5,\,R=1.0,\,s=$WSW featuring $\beta_{opt} = 0.52$, and the model best fitting the MPN (F) $\rho=21,\,R=3.0,\,s=$ASW with $\beta_{opt} = 2.25$.
    }
\end{figure}

In Fig. \ref{fig:si_HM_data} we show the behavior of $\chi^2_b(\beta)$ for the three datasets and the models found to best fit each case. As one can see, we find a minimum of $\chi^2_b(\beta)$ for each class $b$, so that we define the effective $\beta$ of the dataset to be the $\beta_{opt}(b)$ minimizing the curve $\chi^2_b(\beta)$.
All the code to run the analysis can be found in the authors' repository~\cite{github_analysis}.

Let us remember that in the MPN case we have more than one optimal $\beta$ value. This is known to affect the temporal behavior of the system in the asymptotic limit~\cite{Ubaldi_asymptotic_2016}. For example, the growth of the average degree is determined by the minimum value of $\beta$ found in the system. In the previous study, however, we were grouping nodes based on their activity rather than by their entrance time. Here, we apply a different grouping of nodes and, since the model features a single combination of parameters for all the nodes in the system (i.e., we do not assign different $\rho$ and $\nu$ values to the nodes based on some distribution) we focus on the overall $\beta_{opt}$ value of the system (i.e., the one found to be more common among the nodes) when fitting models to empirical data.

\begin{figure}
	\includegraphics[width=.9\textwidth]{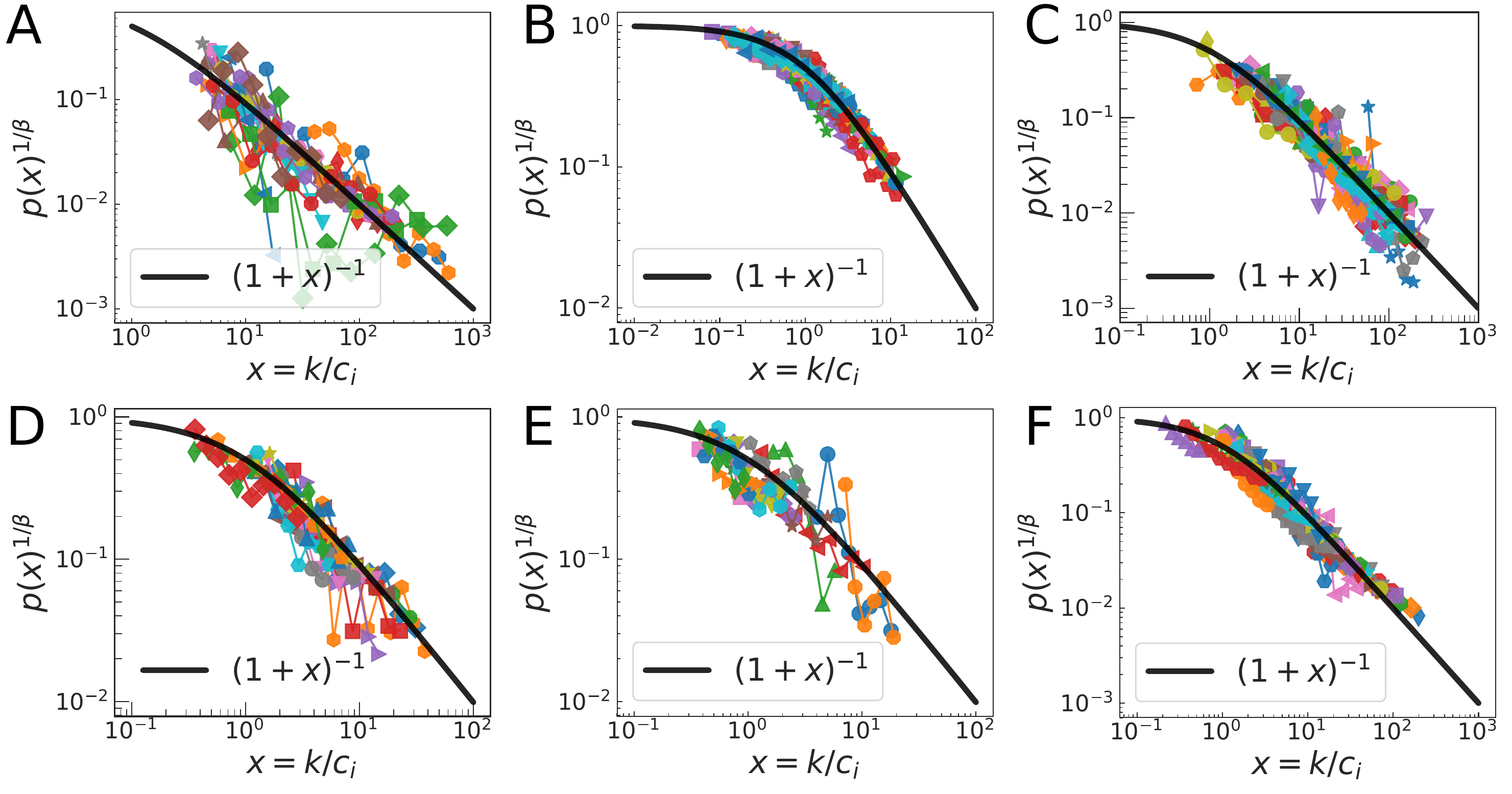}
    \caption{
        \label{fig:si_pk_all}
        Plot of the experimental $p_b(k)$ curves for the (A) APS, (B) MPN, (C) TMN, (D) $\rho=6,\,R=0.4,\,s=$SSW (model fitting APS), (E) the model best fitting the MPN ($\rho=21,\,R=3.0,\,s=$ASW), and, (F) $\rho=5,\,R=1.0,\,s=$WSW (TMN). Here we rescaled $k\to x = k/c_i$, being $c_i$ the strengthening constant of class $i$ and raised $p(x)\to p(x)^{1/\beta}$. The black lines reproduce the theoretical functional form $p(x) = (1+x)^{-1}$.
    }
\end{figure}

In Fig. \ref{fig:si_pk_all} we present the rescaled $p_b(k)$ curves for the APS, TMN and MPN datasets, together with their model counterparts. As one can see, the curves nicely collapse on the reference curve $(1 + k)^{-1}$, highlighting the fact that both in empirical and synthetic data we find the same functional form of the strengthening function $p(k)$.

\subsection{Model fitting}
\label{sub:si_model_fit}

As stated in the main text we run the model at different values of $R$ and $\rho$ for each one of the six sample strategies $s$.
The simulations length measured as the number of steps $T$ depends on the parameter configuration as for small $R$ the system quickly becomes memory consuming (due to the large number of urns entering the system), whereas for large $R$ the weight associated with the most active links (i.e. the number of balls of an ID $i$ in urn $j$ being $e_{ij}$ one of the most active links in the system) quickly overflows the maximum integer value that can be stored in a computer. For these reasons we spanned the $R=(1/2, 2/3, 3/4, 4/5, 1/1, 5/4, 4/3, 7/5, 3/2, 7/4, 9/5, 2/1, 5/2, 3/1, 13/4 7/2, 4/1)$ ratio values and we set:
\begin{itemize}
\item $T=5\cdot10^5$ and $\rho\in[1,15]$ for $R<1$;
\item $T=10^6$ and $\rho\in[1,21]$ for $1 \le R \le 1.5$;
\item $T=10^7$ and $\rho\in[1,30]$ for $R > 1.5$.
\end{itemize}

For each parameters configuration $(\rho, R, s)$ we simulated 10 independent system evolution and for each of them we computed the long-time limit observables being considered in the score $S^d(\rho, R, s)$
\begin{equation}
    S^d(\rho, R, s) = \sum_{i=1}^8{\frac{|o^d_i - \tilde{o_i}(\rho, R, s)|}{\sigma^d_i}},
    \label{eq:si_score}
\end{equation}
where $o^d_i$ and $\sigma^d_i$ are the value and uncertainty on the $i$-th observable of the empirical dataset and $\tilde{o_i}(\rho, R, s)$ the value of the same observable measured in the simulations with configuration $(\rho, R, s)$. The eight selected observables are: 1) the exponent $\gamma$ leading the growth of the number of edges $E(t)\propto t^\gamma$, 2) the optimal $\beta$ measured in the link strengthening function $p(k\to k+1)$, 3) the average clustering coefficient $c$, 4) the exponent leading the growth of the average degree per node class $\av{k(e,t)}\propto t^q$, 4-8) the fractions $OO$, $OC$, $NO$, $NC$ of events allocated toward old/new link insisting or not on a open/closed triangle.
All the code to efficiently run and analyze the simulations can be found on-line~\cite{github_urns}.

\begin{figure}
	\centering
	\includegraphics[width=.65\textwidth]{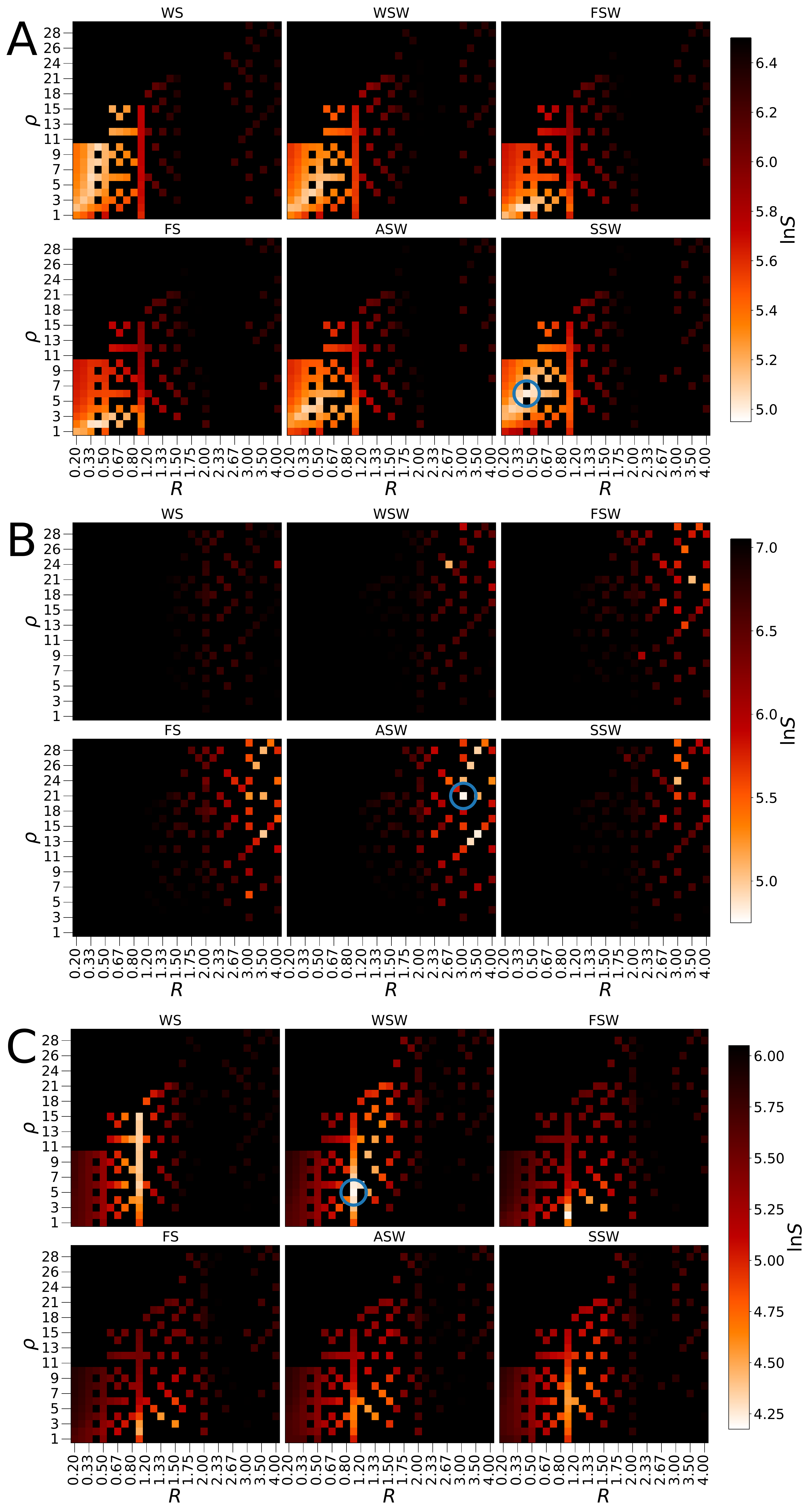}
    \caption{
        \label{fig:si_score}
        Plot of the $\ln{[S^d(\rho, R, s)]}$ score as defined in Eq.~\ref{eq:si_score} for the (A) APS dataset, (B) MPN case, and, (C) TMN data. For each strategy $s$ (subplot) we show the score for all the tested $R$ and $\rho$ values and highlight the best solution found for each case with a blue circle.
    }
\end{figure}

In Fig.~\ref{fig:si_score} we show the behavior of the score for the three datasets. As one can see, each dataset activates a specific region of the $(R,\rho)$ parameter space for each novelties sampling strategy $s$ and, overall, we find a single strategy to better describe each dataset. For example, the APS dataset is clearly well described by the $\rho \lesssim 10$ and $R\lesssim 1$ configurations for all the sampling strategies $s$. However, we observe the best score in the $s=$SSW panel (see Fig.~\ref{fig:si_score}A), specifically in the $(\rho=6,R=0.4,s={\rm SSW})$ point.
On the same page, most of the parameters space badly fits the MPN dataset that turns out to be correctly reproduced only in the $R \sim 3.5$ and $\rho \sim 20$ area. Again, only the $s=$ASW panel displays an overall better score 
whose minimum is found at the $(\rho=21,R=3.0,s={\rm ASW})$ configuration.
Finally, in Fig.~\ref{fig:si_score}C we observe that the TMN dataset is in agreement with the $R\sim 1$ and $\rho\sim 5$ of the $s={\rm WSW}$ strategy, with the best fitting score given by the $(\rho=5,R=1.0,s={\rm WSW})$ parameter combination.

\subsection{Average degree exponent $q$}
\label{sub:si_q_exponent}

\begin{figure}
    \includegraphics[width=0.95\textwidth]{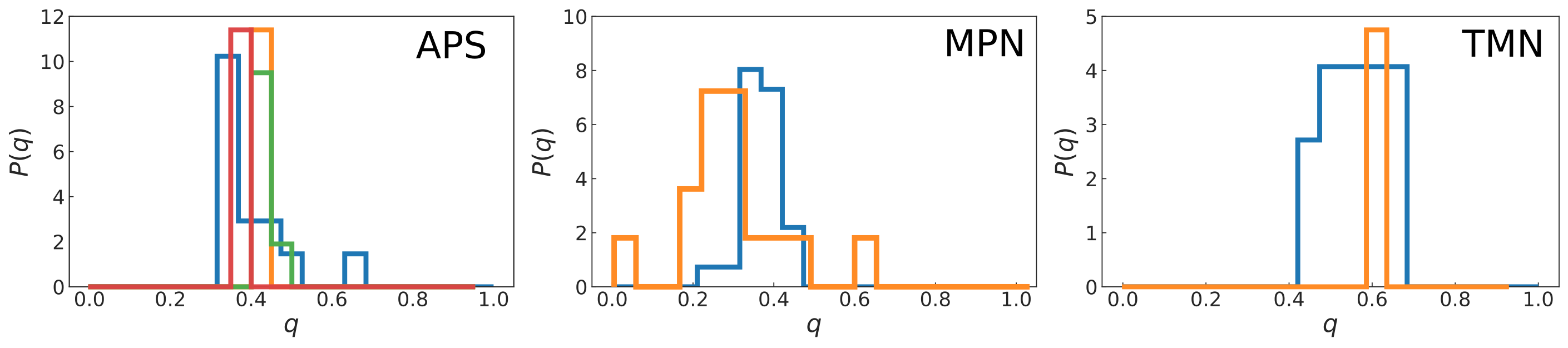}
    \caption{
        \label{fig:si_q_exp} (Left) The $P(q_e)$ exponent distribution in the APS raw data (blue line), the model best fitting it (orange line), the model best fitting the subsampling with one link per paper (green line) and by sampling a number of links equal to the half of authors (red line). (Center) The distribution of the $q_e$ exponent in the MPN dataset (blue line) and in the synthetic case. The same for the TMN in right panel.
    }
\end{figure}

In Fig.~\ref{fig:si_q_exp} we show the $P(q_e)$ distribution of the $q_e$ exponent driving the $\av{k_i(t_e,t)}_{i \in e}\propto (t/t_e)^{q_e}$, for all the $e$ classes found in the system. As one can see the distributions reproduced by the model fit the empirical ones though with lower heterogeneity in some case, as in the TMN one. For the APS data we also show the results coming from different sub-sampling strategies and the corresponding model results (see Section~\ref{sec:si_APS_subsample} below for details).

\subsection{Triggering and entropy measures}
\label{sub:si_entropy}

Another ingredient usually added to the Modified Single Polya Urn model is semantic \emph{triggering}~\cite{tria_dynamics_2014,monechi_waves_2017}. The latter is introduced by reinforcing the fact that each ID in a system composed by a single urn is belongs to a particular semantic group, so that when a certain ID $i$ belonging to a semantic group $g$ gets extracted and put in a sequence, for the next extraction the weight of each balls gets modified accordingly: if a ball features an ID $j$ belonging to the same semantic group $g$ it will have a weight $w_j = 1$, otherwise $w_j = \eta < 1$ (a weight 1 is also assigned to the ball that triggered the entrance of $i$ in the system, i.e., the father of $j$). In this way events in the sequence $\mathcal{S}$ are more likely to be clustered by semantic meaning, in the sense that an ID of the semantic group $g$ will likely trigger one chain of events of IDs belonging to the group $g$. To measure this effect in empirical and synthetic data one can measure the entropy of the appearance of an ID or semantic group $i$ in a sequence of events $\mathcal{S}$. Specifically, if the ID $i$ appeared $k$ times in $\mathcal{S}$, we can compute the local entropy $S_i(k)$ by defining $k$ linearly spaced intervals between $e_i$ (the entrance time, in events, of ID $i$) and the end time $T=|\mathcal{S}|$ of the sequence $\mathcal{S}$. Then we define $f_r$ as the number of occurrences of $i$ in the $r$-th of these $k$ intervals. The entropy of the item then reads:
\begin{equation}
	\label{eq:si_entropy}
	S_i(k) = -\sum_{r=1}^{k} \frac{f_r}{k}\log{\frac{f_r}{k}},
\end{equation}
and takes values between $S_i(k) = 0$ when all the $k$ events are found in a single interval and $S^{\rm max}_i(k)=\log(k)$ (its maximum value) when we observe exactly one event per interval. In the following we then normalize each entropy by its maximum value, so that $S_i(k) = S_i(k)/S^{\rm max}(k)\in [0,1]$.
The entropy measured on the data has then to be compared with the same entropy measured on the shuffled sequence $\mathcal{S}$, where the events' order is randomly changed (and will result in an entropy $S^{rand}_i(k)\ge S_i(k)$).

\begin{figure}
	\centering
    \includegraphics[width=.98\textwidth]{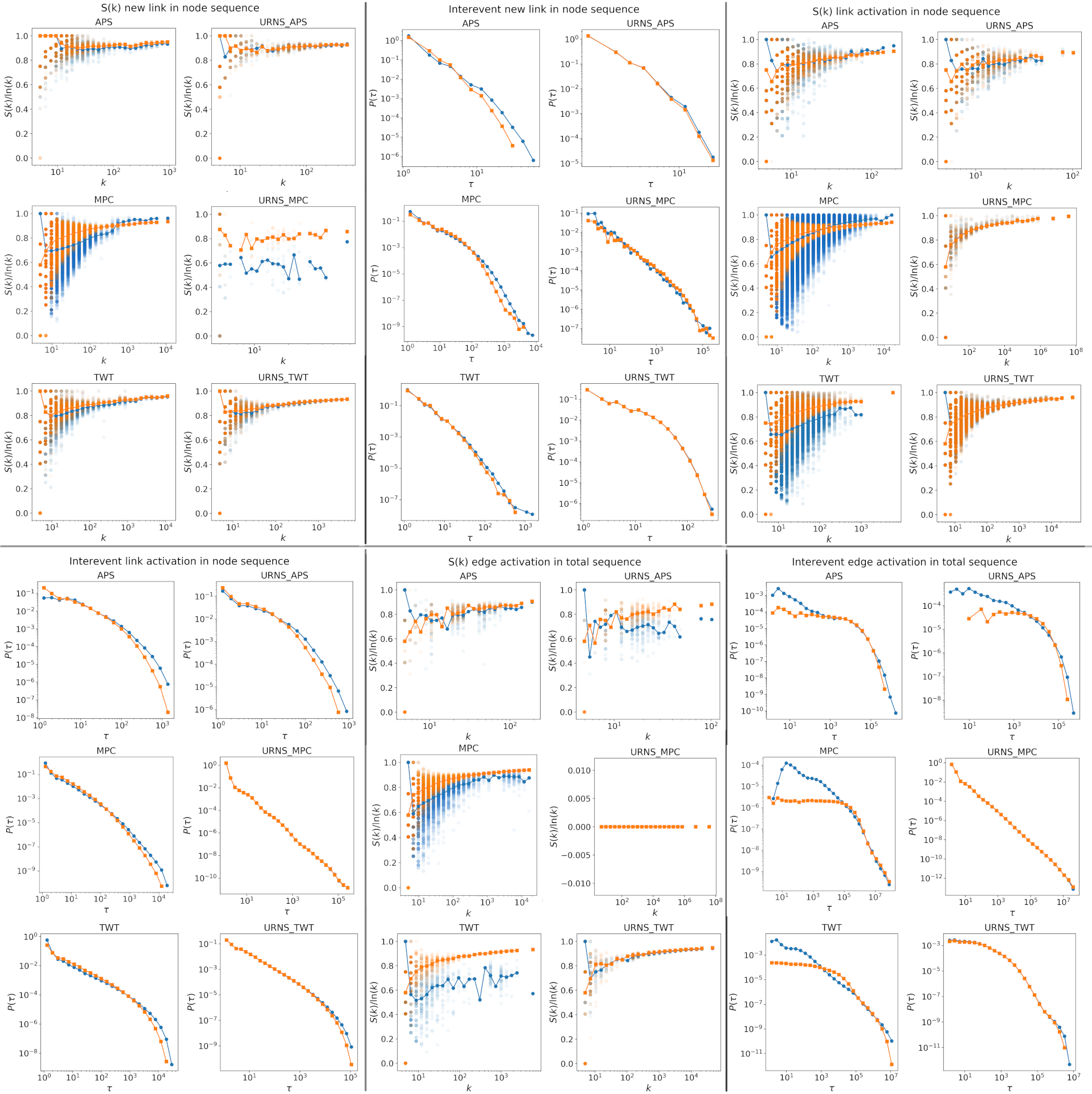}
    \caption{
    	\label{fig:si_entropy}
        We show the entropy $S_i(k)/\ln(k)$ for the (upper left section) new link in node sequence, (upper right) link activation in node sequence, (bottom center) edge activation in total sequence for all the empirical datasets (left column) and the corresponding synthetic data (right column, see subplots titles for the dataset). For each case we also show the inter-event time distribution $P(\tau)$ (top center, bottom left and bottom right sections). We plot the results on the original sequence (blue points) and the randomized sequence (orange points) and the median value of the entropy $S_i(k)$ found for all the nodes with frequency $k$ (solid lines).
    }
\end{figure}

Since in our model we do not have an explicit semantic group and we did not put for simplicity any triggering mechanism we measured the entropy of different possible mechanism that may be subject to triggering. The first is the entropy of the appearance of a new link in the node sequence: for each node we annotate all the events where the node participates and put them in temporal order. Then we put a 0 if the event refers to a link that was already active in the past and 1 if this is the event activating that particular link ---note that we treat the links as undirected so that an event $(i,j)$ is the same as $(j,i)$. We then measure the entropy of the $1$ in this particular sequence and its shuffled version to see whether or not the activation of one link significantly triggers the activation of others in the following events. The results (upper left panel of Fig.~\ref{fig:si_entropy}) show no significant difference between the shuffled and original entropy, highlighting the fact that link activation for a node does not feature triggering.
The same can be seen in the middle top panel of Fig.~\ref{fig:si_entropy} where we plot the inter-event time distribution $P(\tau)$ between new link activation events in the user events sequence. Also in this case we note no difference between the original and the random case. The same behavior is found both in the empirical and synthetic data.

We then applied the same procedure to the sequence of the events of one node $i$ focusing on the activation of a particular link with a neighbor $j$. We then work on the same sequence of events as before putting a 1 if the event is $(i,j)$ or $(j,i)$ and 0 otherwise. This is the entropy of link activation in node sequence of Fig~.\ref{fig:si_entropy} where we observe no signal in the APS case while we observe some signal in the TWT empirical case (but not in the model counterpart). This highlights the fact that in twitter an user interacting with another is more likely to interact again with her and this ingredient is missing at the current stage of our model and could be included in future expansion of the framework.

Finally, we switch to the global (total) sequence of events and check if the activation of one link is clustered in time or not. To this end, for each edge $e=(i,j)$ we cut the main sequence $\mathcal{S}$ between the first and last appearance of an edge $e$ and put a zero if the event activates a link that is not $e$ and 1 otherwise. The results show a moderate signal in the TMN case, showing that interactions between couple of nodes tend to be clustered in time in the global sequence. This weak signal is not significantly observable in the synthetic data since we did not include a triggering process. In all the cases, however, we observe that the synthetic data qualitatively reproduce the tail of the inter-event time distribution $P(\tau)$ (i.e., the number of events occurring between two 1 in the sequences analyzed), so that a further tuning of the model to allow for triggering may improve this agreement in the left part of the distribution.

\subsection{Heaps' law}
\label{sub:si_heaps}

\begin{figure}
	\centering
    \includegraphics[width=.95\textwidth]{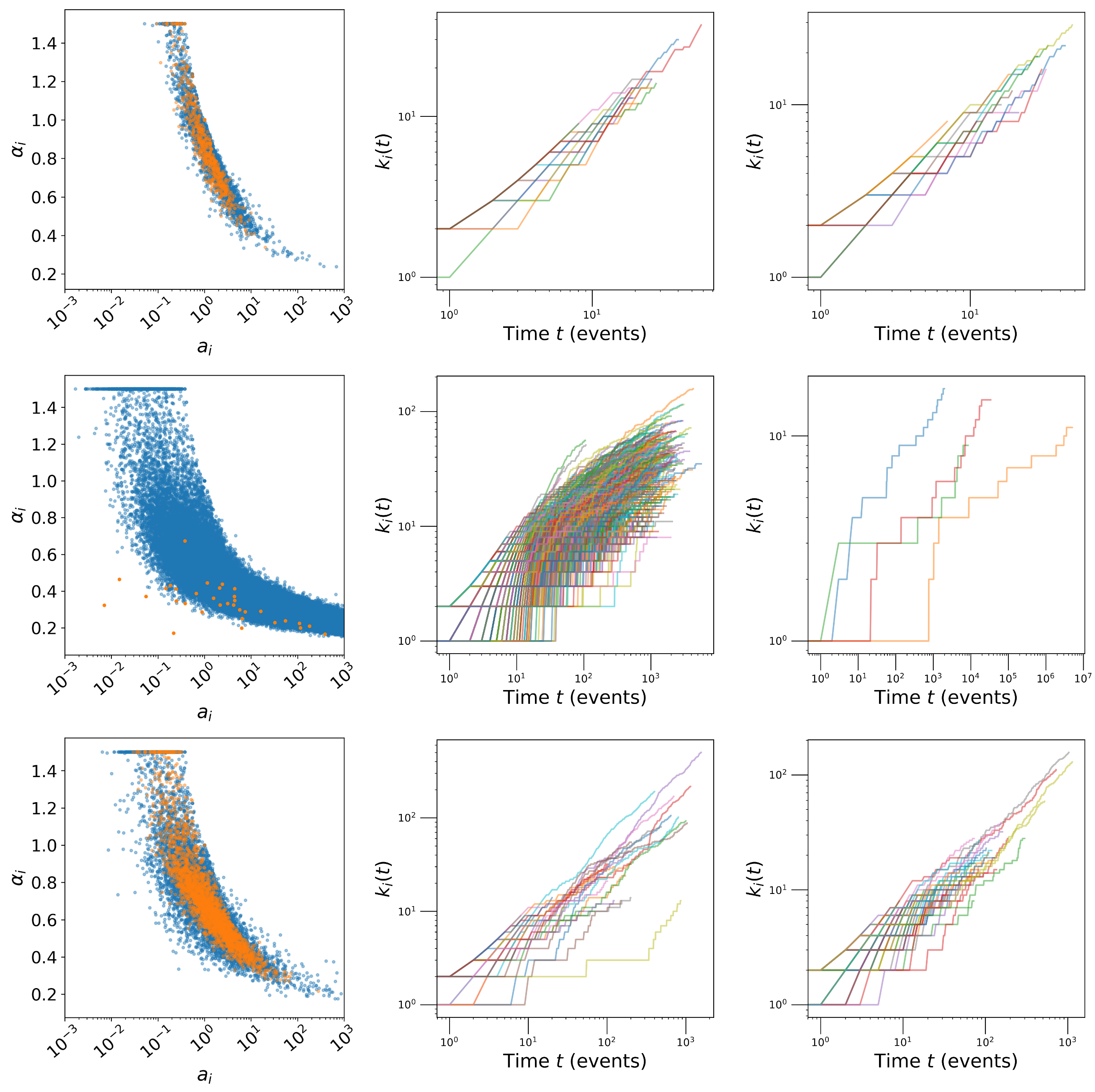}
    \caption{
    	\label{fig:si_heaps}
        We show (first column) the correlations between the Heaps' law parameters $\alpha_i$ and $a_i$, (second column) empirical and (third column) synthetic trajectories of the single node degree in intrinsic time  as found in the (first row) APS, (second row) MPN, and, (third row) TMN data. In the first column empirical data are blue points while synthetic ones are shown as orange markers.
    }
\end{figure}

As stated in the main text we fit the growth of the node degree in intrinsic time (number of events) with an Heaps' law $k_i(t) = (1+a_i t)^{\alpha_i}$ for each node $i$ in the network. The sequence of the node is again obtained considering only the time ordered events $E_i$, $|E_i|=T_i$, in which a node $i$ participates and then, starting from the first event at $t=1$ we assign to each element in the sequence the instantaneous degree $k_i(t)$ for all the position $t=0, \ldots, T_i-1$ (the first element is then $k_i(0)=1$). We then fit the $k_i(t)$ function with the $(1+t/a_i)^{\alpha_i}$.
In this work we adopted this functional form for the Heaps' law to account for an initial transient as the $k_i(t)$ behavior may not start with the power law growth from the beginning but rather after a large number of events (see below and the trajectories in Fig.~\ref{fig:si_heaps}).
Apart from the $P(\alpha)$ distribution shown in the main text we show here the correlations between the $\alpha_i$ and $a_i$ parameters in the empirical data (blue markers in the first column of Fig.~\ref{fig:si_heaps}) that are found to be inversely proportional. So the faster a node accumulates connections (larger $\alpha_i$) the longer it takes for her to converge to the asymptotic behavior (smaller $a_i$). On the other hand it takes a small transient time for nodes slowly exploring their social space to converge to the asymptotic behavior and this usually coincide with a smaller $\alpha_i$.
Surprisingly, we found the same behavior in the data synthetically produced by the model that nicely agree with the real world behavior (orange markers in the first column of Fig.~\ref{fig:si_heaps}).
We show these intuitions by displaying some of these trajectories for empirical (second column of Fig.~\ref{fig:si_heaps}) and synthetic data (third column of Fig.~\ref{fig:si_heaps}).

\subsection{Collective measures}
\label{sub:si_collective}

\begin{figure}
    \centering
    \includegraphics[width=.85\textwidth]{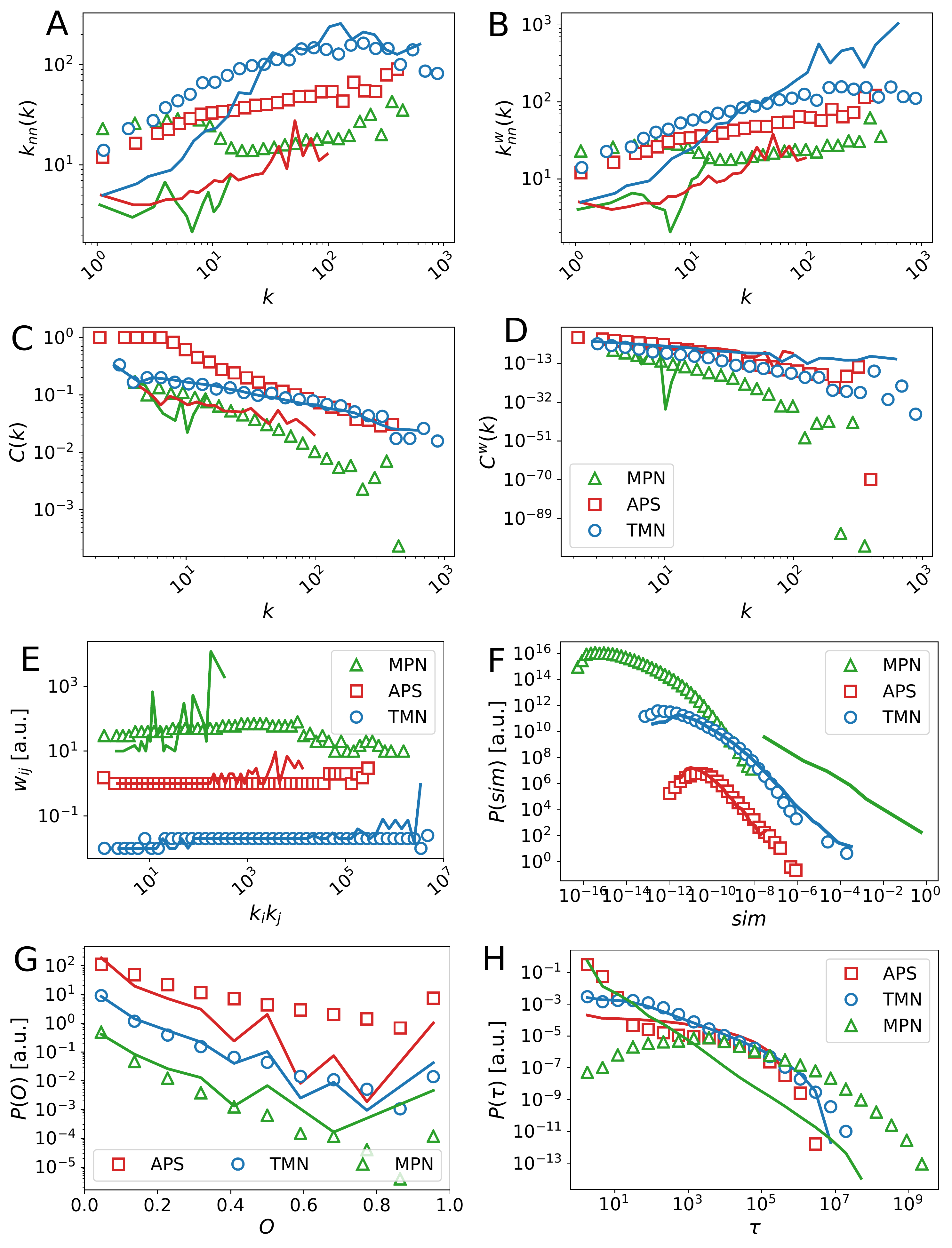}
    \caption{
    \label{fig:si_collective}
        (A) The average nearest neighbors degree $k_{nn}(k)$, (B) the weighted average nearest neighbors degree $k^w_{nn}(k)$, (C) the average local clustering coefficient $C(k)$ for nodes of degree $k$, (D) the average weighted local clustering coefficient $C^w(k)$ for nodes of degree $k$, (E) the $w_{ij}$ link weight as a function of the degree product $k_i k_j$, (F) the similarity distribution $P(sim(i,j))$, (G) the overlap distribution $P(O)$ between two nodes, and, (H) the link activation inter-event time distribution $P(\tau)$. For each observable we show the empirical data of APS (red squares), TMN (blue circles) and MPN (green triangles). The corresponding model results are reported as solid lines featuring the same color of the corresponding markers.
    }
\end{figure}

As a last insight we also performed some more analysis aimed at the exploration of how nodes arrange their outgoing links and their weights among neighbors~\cite{cattuto_collective_2009}.
In addition to the distribution of the link weights $P(w_{ij})$ and the dependence of the average clustering coefficient on the links overlap presented in the main text, we show here some more measures. In particular we check the assortativity of nodes, i.e., the average degree $k_{nn}(k)$ of the nearest neighbors of a node of degree $k$. We measure it both in the unweighted version:
\begin{equation}
	\label{eq:si_kAssortative}
    k_{nn}(k) = \frac{1}{N_k}\sum_i{\delta_{k,k_i} k_{nn,i}},
\end{equation}
where $k_{nn,i} = 1/k_i \sum_{j\sim i} k_j$, $N_k$ is the number of nodes with degree $k$ and $\delta_{k,k_i}$ is the Kronecker delta being 1 if $k_i = k$ and 0 otherwise.
In the same fashion we define the weighted average nearest neighbors degree as:
\begin{equation}
	\label{eq:si_kwAssortative}
    k^w_{nn,i}(k) = \frac{1}{s_i}\sum_{j\sim i}{w_{ij} k_{j}},
\end{equation}
where $s_i$ is the strength (sum of all the weights of the links departing from $i$). We then find the $k^w_{nn}(k) = \av{k^w_{nn,i}(k)}_{(i|k_i = k)}$.
We show the behavior of these functions in Fig.~\ref{fig:si_collective}A-B finding a positive assortativity for all the datasets and for all the corresponding synthetic counterparts. Here, as in the remaining plots, we notice that the MPN synthetic data are small in size compared to the very large dataset of mobile phone calls. Indeed, on our computing facilities we were able to let the model evolve for $10^7$ steps, much less than the billions of events found in the empirical dataset. That is why both the average degree and the inter-event time distributions of the synthetic data have lower cut-offs in the synthetic case. Nevertheless, we observe that the qualitative behavior of the assortativity is reproduced and we find a nice agreement in the APS and TMN cases as well.

We repeat the same procedure to check the correlation between the average clustering coefficient and the node degree.
To this end we measure, for all the nodes with degree $k$:
\begin{equation}
	\label{eq:si_cAssortative}
    C(k) = \frac{1}{N_k}\sum_i{\delta_{k,k_i} c_{i}},
\end{equation}
where $c_i$ is the local clustering coefficient of node $i$.
In the same spirit we define 
\begin{equation}
	\label{eq:si_cwAssortative}
    c^w(i) = \frac{1}{s_i(k_i - 1)} \sum_{j,h\sim i,\,j\sim h} \frac{(w_{ij} w_{ih})}{2},
\end{equation}
that is a weighted clustering coefficient that considers not only the presence of triangles but also how the strength of a node is arranged in the weights insisting on these triangles. Again, we define $C^w(k) = \av{c^w(i)}_{(i|k_i = k)}$.
In this case we observe a disassortative behavior, both in the empirical and synthetic data, where larger degree nodes have lower clustering coefficient values (see Fig.~\ref{fig:si_collective}C-D) and we find, again, a nice agreement between data and the model predictions.

We also check how the product of the degree $k_i k_j$ of the nodes $i$ and $j$ participating in a link correlates with the weight $w_{ij}$ of the link itself. As shown in Fig.~\ref{fig:si_collective}E, the latter seems to weakly depend on the degree product in all the datasets and we recover the same behavior in the model.
On the same page, we also test the $P(sim(i,j))$ distribution of the similarity between vertexes of a single link. The latter is defined as:
\begin{equation}
    \label{eq:si_similarity}
    sim(i,j) = \frac{\sum_k w_{ik} w_{jk}}{\sqrt{\sum_{l} w^2_{il} \sum_{l} w^2_{jl}}},
\end{equation}
and we show the results in Fig.~\ref{fig:si_collective}F. Strikingly, the model also reproduces this property in all the dataset --- in the MPN case we observe larger values of similarity in the model data due to the smaller average degree of the system, nonetheless we qualitatively catch the distribution's tail exponent.

The model also reproduces the overlap distribution $P(O)$, where $O$ is the fraction of common neighbors between two nodes as:
\begin{equation}
    \label{eq:si_overlap}
    O_{ij} = \frac{|\mathcal{V}_i \cap \mathcal{V}_j|}{|\mathcal{V}_i \cup \mathcal{V}_j|},
\end{equation}
where $\mathcal{V}_i$ is the set of neighbors of $i$ and $|\cdot|$ is the cardinality of a set (i.e., the number of elements in it).
As one can see in Fig.~\ref{fig:si_collective}G the overlap distribution is a decreasing function in all the empirical and synthetic datasets and, moreover, the model is able to correctly reproduce also the overall empirical behavior of the $P(O)$.

Finally, in Fig.~\ref{fig:si_collective}H we show the inter-event time distribution between links activation in empirical time, i.e., the number of events between the activation of a link $e_{ij}$. We find a very nice agreement in the APS and TMN cases, whereas in the MPN case the empirical distribution is more heterogeneous than the synthetic one due to the different order of magnitude of the number of events in the data. However, we qualitatively reproduce the right tail exponent of the distribution.

\section{APS subsampling} 
\label{sec:si_APS_subsample}

In this section we investigate whether or not the different behavior of the empirical and synthetic APS data stems from the "clique" nature of the dataset. Indeed, in the APS data we transform each paper published by $n$ authors in a sequence of $E=n(n-1)$ events with all the possible links between all the ordered couples of co-authors. It is evident that, at the current stage, the model cannot replicate this kind of dynamics and quantity of contacts.
It is then reasonable to check what is happening if we consider only a subset of the possible interactions that each paper brings into the system. A possible way to do so is to sample $l$ links over the $E$ possible links for each paper to be inserted in the total sequence $\mathcal{S}$. In the following we present the results obtained by performing 10 sampling procedures sampling 1, 2, 3, 4, 5, and, 10 links per paper. In every case we consider up to $E$ links for each paper (i.e., we do not sample twice a link if we have to sample 10 links from a paper with, say, 3 authors).

\begin{figure}
    \centering
    \includegraphics[width=0.7\textwidth]{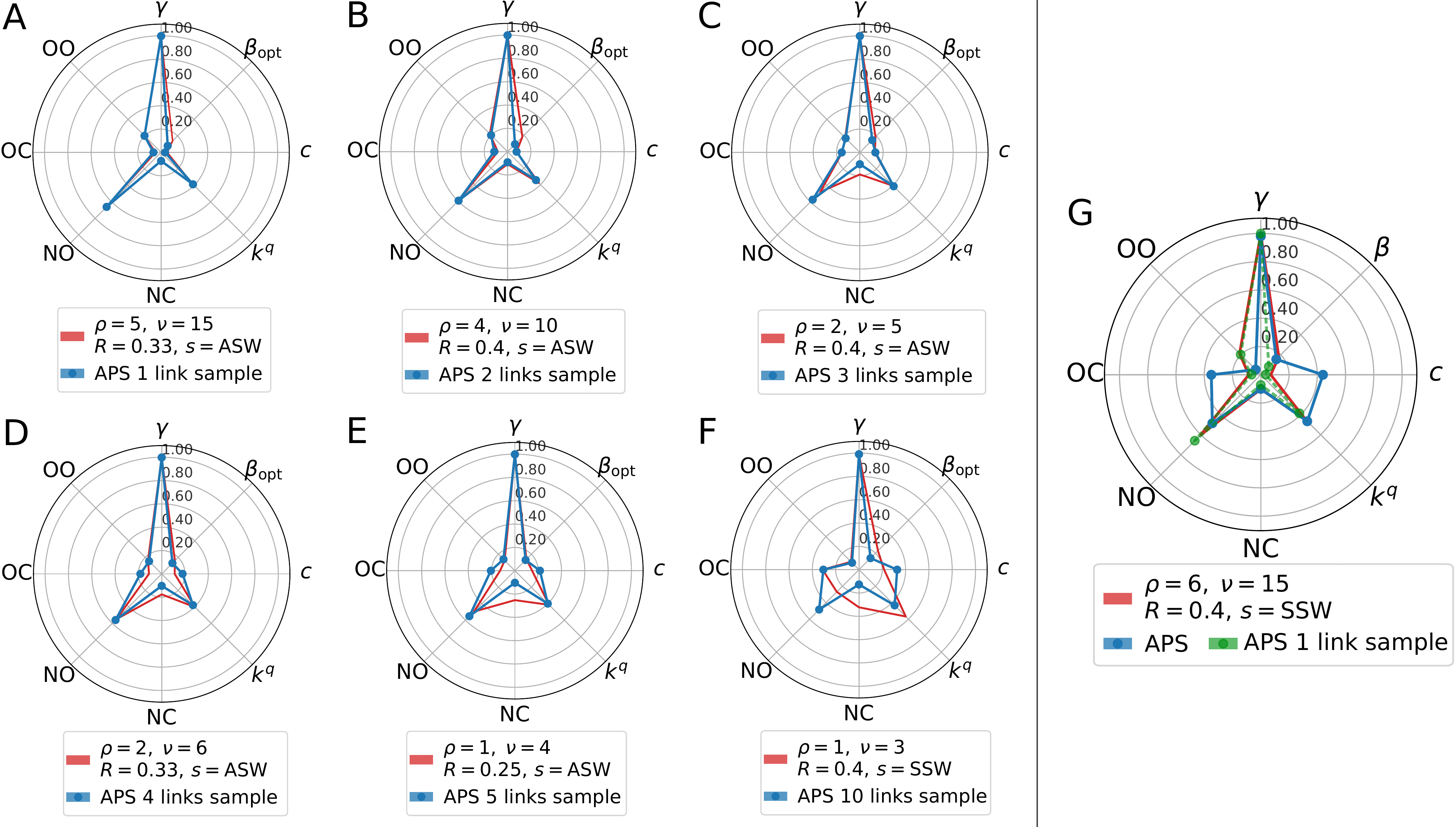}
    \caption{
        \label{fig:si_radars_APS_samples} The radar plots for the APS sub-samples with 1 (A), 2 (B), 3 (C), 4(D), 5 (E), and, 10 (F) links. For each dataset we plot the empirical results (blue line) and the best model fitting data (red line). In (G) we show the empirical results of the entire dataset (blue line) compared with the best fitting model (red line) and the results of the 1-link sub-sample dataset (green dotted line).
    }
\end{figure}

In Fig.~\ref{fig:si_radars_APS_samples} we show the radar plots showing the temporal and topological properties included in the score computation of Eq.~(\ref{eq:si_score}) for the networks obtained considering $l=[1,2,3,4,5,10]$ sub-sampled links. We see that the system is able to reproduce the main characteristics of all the sub-samples up to about $l\sim5$. For $l=10$ the system cannot replicate both the increasing clustering coefficient $c$ and the number of events insisting on new and old links insisting or not on a triangle at once. 
Another interesting hint is given in Table~(\ref{tab:si_sampling}) where we present the numerical dependence of the measured observables and the model parameters as a function of the number of links sampled $l$.
Specifically, we see that the system is always compatible with a $R\sim 1/3$ ratio between reinforcement and innovation. Moreover, the $q$ exponent leading the average degree growth is stable at $q\sim 0.4$ as we expect the sub-sampling of edges not to change the growth rate of the network but rather its degree correlation properties. Indeed, we observe a steady increase of both the clustering coefficient $c$ and of the old links insisting on closed triangle, whereas we find a decrease of the number of events toward both new and old links insisting on new triangles, as one can reasonably expect.

Finally, we observe that the optimal strategy $s$ to switch from an asymmetric one for low $l$ (ASW, i.e., asymmetric sliding window) to a symmetric one when $l=10$ and for the total dataset. This finding confirms that the score of Eq.~(\ref{eq:si_score}) is able to give valuable insights on the dynamics ongoing globally and locally in the network, as it is able to detect the increased number of reciprocal links in the dataset as $l$ increases, i.e., as the network converges to the real one, that is entirely composed of cliques.

\begin{figure}
    \centering
    \includegraphics[width=0.75\textwidth]{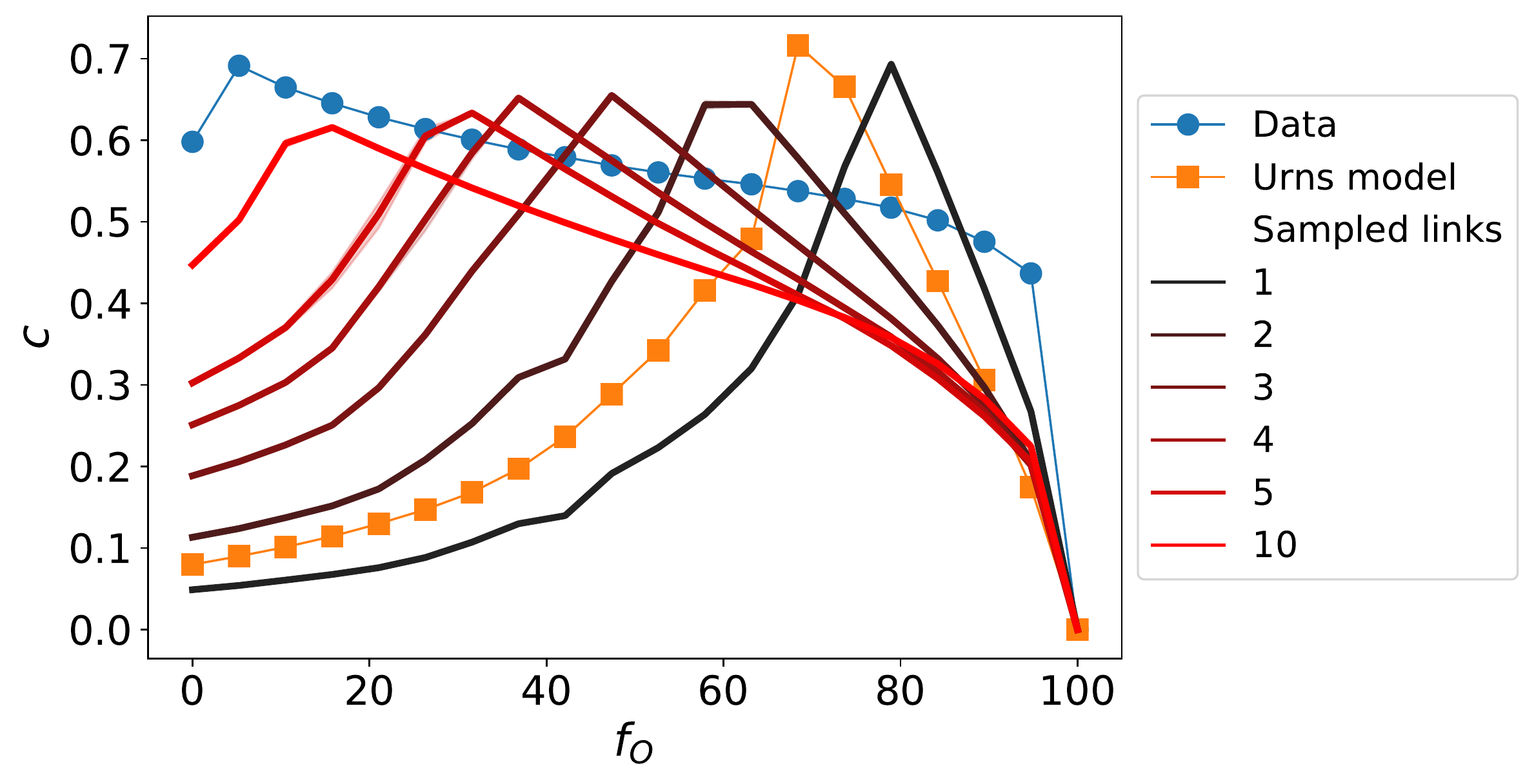}
    \caption{
        \label{fig:si_cluster_APS_samples} The curves of the average clustering coefficient as a function of the fraction of removed edges (sorted by the overlap of the nodes they are insisting on) for the empirical dataset (blue dotted line) the urn model best fitting it ($\rho=6,\,\nu=15,\,s=SSW$, orange squared line) and the curves measured in 10 samplings repetitions taking from 1 link (black solid line) to 10 links (red line, intermediate values in tones of red, see legend for values). We also show the curve of the model best fitting the 1-link subsample data  ($\rho=5,\,\nu=15,\,s=ASW$, green squared line) for comparison.
    }
\end{figure}

To conclude our analysis, in Fig.~\ref{fig:si_cluster_APS_samples} we show that the same holds for the degree correlations in the total and sampled networks. Indeed, as long as we remove links from the dataset, the $c(f_O)$ curve converges to the one found in the best synthetic model fitting the APS dataset ($\rho=6,\,\nu=15,\,s=SSW$). Indeed, when lowering $l$ we remove the least active edges in the system (that are less likely to be selected in the sub-sample), thus only retaining the strong links within the actual communities. As a consequence, less overlapping nodes have a smaller amount of triangles in common with respect to the previous case and the overall average clustering coefficient is lower. On the other hand, the communities' core still feature high clustering coefficient $c\sim0.7$ but they are now revealed only when we arrive at the $f_O\sim 80\%$ value (for $l=1$).

\begin{table}
    \centering
    \begin{tabular}{|c|c|c|c|c|c|c|c|c|c|c|c|c|c|c|}
        \hline
        Links $l$  & Case & $\rho$  &   $\nu$  & $R$ &   $s$ & $\gamma$ & $\beta_{\rm opt}$ & $c$ & $q$ & $NC$ & $NO$ & $OC$ & $OO$\\
        \hline
        \hline
        1    & Data        & -  &   -         & -    &   -   & 0.998 & 0.080 & 0.034 & 0.387 & 0.073 & 0.661 & 0.065 & 0.201 \\
             & Model       & 5  &   15        & 0.33 &   ASW & 1.00  & 0.140 & 0.053 & 0.379 & 0.078 & 0.678 & 0.048 & 0.197 \\
        \hline
        2    & Data        & -  &   -         & -    &   -   & 0.998 & 0.090 & 0.077 & 0.345 & 0.091 & 0.594 & 0.114 & 0.200 \\
             & Model       & 4  &   10        & 0.25 &   ASW & 1.00  & 0.180 & 0.077 & 0.380 & 0.107 & 0.594 & 0.083 & 0.215 \\
        \hline
        3    & Data        & -  &   -         & -    &   -   & 0.998 & 0.150 & 0.133 & 0.410 & 0.101 & 0.573 & 0.155 & 0.172 \\
             & Model       & 2  &   5         & 0.40 &   ASW & 1.00  & 0.190 & 0.129 & 0.405 & 0.189 & 0.481 & 0.152 & 0.179 \\
        \hline
        4    & Data        & -  &   -         & -    &   -   & 0.998 & 0.130 & 0.179 & 0.379 & 0.103 & 0.560 & 0.183 & 0.153 \\
             & Model       & 2  &   6         & 0.33 &   ASW & 1.00  & 0.160 & 0.111 & 0.403 & 0.176 & 0.549 & 0.109 & 0.166 \\
        \hline
        5    & Data        & -  &   -         & -    &   -   & 0.998 & 0.130 & 0.215 & 0.399 & 0.103 & 0.552 & 0.206 & 0.138 \\
             & Model       & 1  &   4         & 0.25 &   ASW & 1.00  & 0.140 & 0.148 & 0.414 & 0.252 & 0.496 & 0.127 & 0.124 \\
        \hline
        10   & Data        & -  &   -         & -    &   -   & 0.990 & 0.140 & 0.328 & 0.433 & 0.127 & 0.483 & 0.306 & 0.084 \\
             & Model       & 1  &   3         & 0.33 &   SSW & 1.00  & 0.230 & 0.216 & 0.568 & 0.323 & 0.269 & 0.312 & 0.096 \\
        \hline
        \hline
        All  & Data        & -  &   -         & -    &   -   & 0.980 & 0.155 & 0.440 & 0.465 & 0.100 & 0.485 & 0.350 & 0.052 \\
             & Model       & 6  &   15        & 0.40 &   SSW & 0.999 & 0.180 & 0.071 & 0.452 & 0.102 & 0.600 & 0.085 & 0.212 \\
        \hline
    \end{tabular}
    \caption{
        \label{tab:si_sampling} The table resuming the results as found in the different sub-samples and in the total dataset. For each sub-sample (number of links $l$), we report the corresponding best fit parameters $\rho$, $\nu$ and $s$ (together with the ratio $R=\rho/\nu$) and the eight observables included in the score defined in Eq.~(\ref{eq:si_score}). For each number of links sampled we show the synthetic data observables (first row of the block) as well as the parameters and observables of the best fitting model found for that sample (second row of the block).
    }
\end{table}

\bibliographystyle{unsrt}
\bibliography{bibliography}